\documentclass[a4paper,fleqn,usenatbib]{mnras}

\usepackage{newtxtext,newtxmath}

\usepackage[T1]{fontenc}
\usepackage{ae,aecompl}

\usepackage{graphicx}	
\usepackage{amsmath}	
\usepackage{wasysym}

\def\chisqr{\hbox{$\chi^2_{\rm r}$}}
\def\msun{\hbox{${\rm M}_{\odot}$}}
\def\mjup{\hbox{${\rm M}_{\jupiter}$}}

\def\mspy{\hbox{${\rm M}_{\odot}$\,yr$^{-1}$}}
\def\rsun{\hbox{${\rm R}_{\odot}$}}
\def\lsun{\hbox{${\rm L}_{\odot}$}}
\def\rcor{\hbox{$r_{\rm cor}$}}
\def\rmag{\hbox{$r_{\rm mag}$}}
\def\mstar{\hbox{$M_{\star}$}}
\def\rstar{\hbox{$R_{\star}$}}
\def\lstar{\hbox{$L_{\star}$}}
\def\teff{\hbox{$T_{\rm eff}$}}
\def\logg{\hbox{$\log g$}}

\def\kms{\hbox{km\,s$^{-1}$}}

\def\vsini{\hbox{$v \sin i$}}

\def\mic{\hbox{$\mu$m}}

\def\emr{}
\def\Bl{\hbox{$B_{\rm \ell}$}}
\def\Bd{\hbox{$B_{\rm d}$}}

\def\degr{\hbox{$^\circ$}}

\def\Mdot{\hbox{$\dot{M}$}}

\def\Porb{\hbox{$P_{\rm orb}$}}
\def\Prot{\hbox{$P_{\rm rot}$}}

\newcommand{\hei}{\hbox{He$\;${\sc i}}}

\newcommand{\pab}{\hbox{Pa${\beta}$}}
\newcommand{\brg}{\hbox{Br${\gamma}$}}

\title[Binarity, magnetic field and accretion of V347~Aur]{SPIRou monitoring of the protostar V347~Aur: binarity, magnetic fields, pulsed dynamo and accretion} 
\author[J.-F.~Donati et al.]{J.-F.~Donati$^{1}$\thanks{E-mail: jean-francois.donati@irap.omp.eu},
           P.I.~Cristofari$^{2}$, A.~Carmona$^{3}$, K.~Grankin$^{4}$ 
\vspace{1mm}
\\ 
$^1$ Universit\'e de Toulouse, CNRS, IRAP, 14 avenue Belin, 31400 Toulouse, France \\ 
$^2$ Center for Astrophysics, Harvard \& Smithsonian, 60 Garden street, Cambridge, MA 02138, United States \\ 
$^3$ Universit\'e Grenoble Alpes, CNRS, IPAG, 38000 Grenoble, France  \\
$^4$ Crimean Astrophysical Observatory, Nauchny, Crimea 298409  
}

\date{Accepted 2024 September 02. Received 2024 August 09; in original form 2024 July 06.} 

\pubyear{2023}

\begin{document}

\label{firstpage}
\pagerange{\pageref{firstpage}--\pageref{lastpage}}
\maketitle

\begin{abstract}
We present in this paper an analysis of near-infrared observations of the 0.3~\msun\ protostar V347~Aur collected with the SPIRou high-resolution spectropolarimeter and 
velocimeter at the 3.6-m Canada-France-Hawaii Telescope from October 2019 to April 2023.  From a set of 79 unpolarized and circularly polarized spectra of V347~Aur to 
which we applied Least-Squares Deconvolution (LSD), we derived radial velocities and longitudinal fields, along with their temporal variations over our monitoring 
campaign of 1258~d.  Our data show that V347~Aur is an eccentric binary system with an orbital period $154.6\pm0.7$~d, experiencing strong to extreme accretion events near 
periastron.  The companion is a $29.0\pm1.6$~\mjup\ brown dwarf, a rare member of the brown dwarf desert of close companions around M dwarfs.  
We detect weak longitudinal fields ($<$100~G) at the surface of V347~Aur, significantly weaker than those of more evolved prototypical T~Tauri stars.  These fields show small-amplitude 
rotational modulation, indicating a mainly axisymmetric parent large-scale magnetic topology, and larger fluctuations at half the orbital period, suggesting that what we dub {\emr a ``pulsed 
dynamo'' triggered by orbital motion and pulsed accretion} operates in V347~Aur.  Applying Zeeman-Doppler imaging to our circularly polarised LSD profiles, we find that the large-scale field 
of V347~Aur is mainly toroidal for most of our observations, with the toroidal component switching sign near periastron and apoastron.  The weak large-scale dipole ($\simeq30$~G) is not able 
to disrupt the disc beyond 1.3~\rstar\ even {\emr at the lowest accretion rates}, implying longitudinally distributed (rather than localized) accretion at the surface of the protostar.  
\end{abstract}




\section{Introduction}
\label{sec:int}

Stars are born from giant filamentary molecular clouds locally collapsing under the effect of gravitation and turbulence, forming protostellar embryos 
feeding from their accretion discs and known as Class-0 then Class-I objects, with the youngest (Class-0) ones having yet accreted no more than a small fraction of 
their final mass.  Embedded within dusty envelopes, protostars are opaque at optical wavelengths, with Class-0 ones mostly detectable at millimeter and radio wavelengths, 
{\emr sometimes in the near-infrared \citep[nIR, e.g.,][]{Greene18,LeGouellec24}}, while the more evolved Class-I objects are visible at nIR wavelengths and redwards \citep[e.g.,][]{Andre09}.  
Class~I objects subsequently evolve into pre-main sequence (PMS) stars (T~Tauri stars for low-mass objects), lacking a dense circumstellar envelope and thereby observable at optical 
wavelengths.  They are called Class-II objects when they still accrete from their discs and Class-III objects once accretion no longer occurs at the surface of the star.  
Planetary systems are presumably born simultaneously with the central star, from the accretion discs formed during the cloud collapse \citep[e.g.,][]{Hennebelle20}.  

Magnetic fields play a key role at all stages and scales in this process, within dense cores, accretion discs and associated outflows in the earliest phases,  
and within protostars and PMS stars once these gathered most of the surrounding mass and become the main driver of the formation 
process.  In particular, magnetic fields are likely responsible for reducing the rate of star formation, for causing magnetic braking that impacts the outcome 
of the star formation process, for generating outflows and jets evacuating most of the initial angular momentum, for regulating the rotation rates of protostars 
and PMS stars by coupling them to the inner regions of their accretion discs, and for affecting planetary formation and migration within protoplanetary discs 
\citep[e.g.,][]{Romanova02,Bouvier07,Hennebelle08a,Hennebelle08b,Zanni13,Bouvier14,Blinova16,Hennebelle19}.  

Whereas magnetic fields of star-forming cores and protostellar discs of Class-0 and Class-I objects are being extensively characterized at millimeter wavelengths 
\citep[e.g.,][]{Maury22}, those of accreting and non-accreting T~Tauri stars are scrutinized in the optical and nIR domains using high-resolution spectroscopy 
and spectropolarimetry \citep[e.g.,][]{Johns07,Donati20b,Lopez-Valdivia21,Finociety21,Finociety23,Donati24}, focussing in particular on the dynamo processes 
amplifying such fields and on the magnetospheric accretion processes taking place between the stars and their inner discs.  However, very few studies exist 
regarding magnetic fields of class-I propostars and the way they accrete mass from their accretion discs \citep{Johns09,Flores19,Flores24}.  Yet this is a 
critical piece of the puzzle for us to bridge the gap between both sets of studies, and ultimately merge all observational constraints about the role of magnetic 
fields at all temporal stages and spatial scales of the star and planet formation process.  

In this aim, the young protostar V347~Aur is a particularly interesting object.  With a spectral index $\alpha$ (describing the slope of the spectral energy 
distribution between near- and mid-infrared wavelengths) close to 0 \citep{Connelley10}, V347~Aur is classified as a flat-spectrum source, in transition 
between Class-I and Class-II formation stages \citep{Greene94}.  Relatively bright in the nIR, V347~Aur was recently characterized through high-resolution 
spectroscopy \citep{Flores19,Flores24}, showing in particular that it hosts kG small-scale magnetic fields at its surface (from the broadening and 
intensification of spectral features sensitive to magnetic fields).  V347~Aur is also known to undergo frequent and regular accretion outbursts 
\citep{Dahm20} possibly caused by the presence of an unseen companion, making it even more suited for an in-depth monitoring study of its spectral 
properties.  

We thus embarked on a multi-season program using SPIRou, the nIR high-resolution cryogenic spectropolarimeter / velocimeter at the Canada-France-Hawaii Telescope 
(CFHT), an ideal instrument to monitor the spectral properties of V347~Aur, and in particular its large- and small-scale magnetic field and its velocimetric 
behaviour.  The present paper describes a monitoring effort of V347~Aur carried out over 3.5~yr, from late 2019 to early 2023.  We start by summarizing the main 
properties of V347~Aur in Sec.~\ref{sec:par}, then present our SPIRou observations of V347~Aur in Sec.~\ref{sec:obs}, outline our spectropolarimetric and velocimetric 
results in Secs.~\ref{sec:bl} and \ref{sec:rvs}, and describe our magnetic modeling of V347~Aur using Zeeman-Doppler imaging (ZDI) in Sec.~\ref{sec:zdi}.  We finally 
describe in Sec.~\ref{sec:eml} the accretion properties of V347~Aur as derived from the variability of photospheric and accretion lines, before summarizing and 
discussing our results in Sec.~\ref{sec:dis}.

\section{The protostar V347~Aur}
\label{sec:par}

V347~Aurigae (HBC~428, IRAS~04530+5126, 2MASS~J04565702+5130509) is a highly variable young star located $206.6\pm2.3$~pc away from the Sun 
\citep{Gaia23} in the Taurus-Auriga complex, relatively isolated from known star-forming regions.  Being still embedded within its natal 
molecular cloud L1438, V347~Aur interacts with the surrounding nebular material and is associated with a small reflection nebula.  
With an emission line spectrum and a location in a color-magnitude diagram that are characteristic of an extremely young object, 
it was first classified as a Class-I protostar of spectral type M2 suffering significant extinction \citep{Cohen78}, then shown to be a 
flat spectrum source in transition towards a Class-II T~Tauri star \citep{Connelley10}.    

From low- and high-resolution nIR spectra, V347~Aur was confirmed to be an M2 star \citep[with spectral type varying from M1 to M3,][]{Connelley10,
Dahm20}, corresponding to an average effective temperature $\teff\simeq3500$~K \citep{Pecaut13}.  By fitting recent high-resolution nIR spectra of 
V347~Aur, \citet{Flores19} finds a lower \teff\ of $3230\pm100$~K and a logarithmic gravity of $\logg=3.25\pm0.15$ (in cgs units), 
and reports the presence of a small-scale magnetic field of average strength $1.36\pm0.06$~kG at the surface of the star, broadening and intensifying 
the profiles of atomic lines \citep[with all measurements confirmed within error bars in their new study,][]{Flores24}.  {\emr Our own measurements from 
SPIRou spectra, yielding similar results ($\teff=3340\pm50$~K and $\logg=3.30\pm0.10$, see Sec.~\ref{sec:obs}), further confirm those of 
\citet{Flores19}.  We use a combination of both, i.e., $\teff=3300\pm50$~K and $\logg=3.30\pm0.10$, in the following sections. } 

Adjusting both \teff\ and \logg\ to the evolutionary tracks of \citet{Baraffe15} for low-mass stars yields a mass of $\mstar=0.25\pm0.03$~\msun, a radius 
of $\rstar=1.9\pm0.3$~\rsun\ and an age of $\simeq$0.8~Myr, with similar results when using the non-magnetic tracks of \citet{Feiden16}, both sets of 
models being known to underestimate masses by $\simeq$10~per cent for stars with $\mstar<0.6$~\msun\ \citep{Braun21}.  Comparing now with the magnetic tracks of \citet{Feiden16} 
yields $\mstar=0.45\pm0.06$~\msun, $\rstar=2.4\pm0.3$~\rsun\ and an age of $\simeq$0.8~Myr.  Although presumably more adapted to a magnetic star like 
V347~Aur, these models are however reported to overestimate masses by $\simeq$30~per cent for the same mass range \citep{Braun21}.  Correcting for these under and 
overestimates, and taking the average of the corrected values yields $\mstar=0.33\pm0.05$~\msun, $\rstar=2.2\pm0.3$~\rsun\ (hence a logarithmic bolometric 
luminosity relative to the Sun $\log(\lstar/\lsun)=-0.3\pm0.2$) and an age of $\simeq$0.8~Myr for V347~Aur, a set of parameters roughly consistent with that 
of \citet{Dahm20} and which we use in the rest of our study.  


Our measurements of the rotation period ($\Prot=4.12\pm0.03$~d, see Sec.~\ref{sec:bl}) and of the line-of-sight projected equatorial rotation 
velocity \citep[$\vsini=12.0\pm0.5$~\kms, see Sec.~\ref{sec:zdi}, in good agreement with][]{Flores19} of V347~Aur independently yield 
$\rstar \sin i=0.98\pm0.05$~\rsun, implying an inclination angle $i$ of the rotation axis of V347~Aur to the line of sight of $i=26\pm5\degr$.  
Assuming an accretion disc whose rotation axis is aligned with that of the star, it implies a system geometrical orientation that is only 
$\simeq$26\degr\ away from being face-on.  Little is known about the disc itself beyond an old radio measurement of the 1.3~mm dust continuum 
with the 30-m IRAM interferometer, which derived a small disk mass of only 0.0014~\msun\ \citep{Osterloh95}.  
We list the main stellar parameters of V347~Aur in Table~\ref{tab:par} and adopt values of $i=26$\degr, $\mstar=0.33~\msun$ and $\rstar=2.2~\rsun$ in the following.  

V347~Aur was reported to undergo regular accretion outbursts generating photometric pulses with a period of about 155~d, featuring sharp (30~d) rises followed 
by slower (60~d) subsequent decays, and with brightness fluctuations varying from cycle to cycle by more than an order of magnitude \citep{Dahm20}.  V347~Aur is also observed 
to get bluer during the bursts and redder afterwards \citep{Dahm20}.  One obvious explanation for this phenomenon could be that V347~Aur is the most massive star of an 
eccentric binary where both stars interact during periastron passage generating enhanced accretion, as in, e.g., the close T~Tauri binary 
DQ~Tau \citep{Mathieu97,Fiorellino22,Pouilly23,Pouilly24} or the more massive Herbig Ae + T~Tauri system HD~104237 \citep{Dunhill15, Jarvinen19}.  
\citet{Dahm20} to conclude 
that V347~Aur is not a spectroscopic binary like DQ~Tau or HD~104237.  A second option put forward by \citet{Dahm20} on the basis of the large amplitude of the observed photometric 
variations (by factors up to tens in brightness) is that V347~Aur experiences instabilities in its accretion disc, possibly triggered by a low-mass companion further out 
in the disc.  The reported periodicity of 155~d would imply the companion to be located at an average distance of about 0.4~au from V347~Aur.  We show further 
(see Sec.~\ref{sec:rvs}) that V347~Aur is indeed a spectroscopic binary with a low-mass companion, featuring RV variations of about 2~\kms\ peak-to-peak.  

\begin{table}
\caption[]{Parameters of V347~Aur used in / derived from our study} 
\scalebox{0.95}{\hspace{-4mm}
\begin{tabular}{ccc}
\hline
distance (pc)        & $206.6\pm2.3$   & \citet{Gaia23} \\
$\log(\lstar/\lsun)$ & $-0.3\pm0.2$    & \\ 
\teff\ (K)           & $3300\pm50$     & \\
\logg\ (dex)         & $3.30\pm0.10$   & \\ 
\mstar\ (\msun)      & $0.33\pm0.05$   & \citet{Baraffe15, Feiden16} \\
\rstar\ (\rsun)      & $2.2\pm0.3$     & \citet{Baraffe15, Feiden16} \\
age (Myr)            & $\simeq$0.8     & \citet{Baraffe15, Feiden16} \\ 
\Prot\ (d)           & $4.12$          & period used to phase data \\ 
\Prot\ (d)           & $4.12\pm0.03$   & from RV data \\ 
\vsini\ (\kms)       & $12.0\pm0.5$    & from ZDI modeling \\ 
$\rstar \sin i$ (\rsun)& $0.98\pm0.05$ & from \rstar\ and \vsini \\ 
$i$ (\degr)          & $26\pm5$        & from \rstar\ and $\rstar \sin i$ \\ 
\rcor\ (au)          & $0.035\pm0.002$ & from \mstar\ and \Prot \\ 
\rcor\ (\rstar)      & $3.4\pm0.6$     & from \rcor\ (au) and \rstar \\ 
$\log\Mdot$ (\mspy)  & $-9.0$ to $-6.2$& from \pab\ and \brg \\ 
\hline
\end{tabular}}
\label{tab:par}
\end{table}

\section{SPIRou observations of V347~Aur}
\label{sec:obs}

We observed V347~Aur in 4 consecutive seasons with the SPIRou nIR spectropolarimeter / high-precision velocimeter \citep{Donati20} at CFHT, within the 
SPIRou Legacy Survey (SLS) in 2019-2020, 2020-2021 and early 2022, then shortly within the SPICE Large Programme in 2022-2023.  SPIRou collects unpolarized and 
polarized stellar spectra, covering a wavelength interval of 0.95--2.50~\mic\ at a resolving power of 70\,000 in a single exposure.  For the present 
study, we only recorded circularly polarized (Stokes $V$) and unpolarized (Stokes $I$) spectra.  As in previous studies, polarization observations consist of 
sequences of 4 sub-exposures, with each sub-exposure associated with a different azimuth of the Fresnel rhomb retarders of the SPIRou polarimetric unit 
in order to remove systematics in polarization spectra \citep[to first order, see, e.g.,][]{Donati97b}.  Each recorded sequence yields one Stokes $I$ 
and one Stokes $V$ spectrum, along with a null polarization check (called $N$) used to diagnose potential instrumental or data reduction issues.  

A total of 89 polarization sequences were recorded for V347~Aur over 4 main seasons, 18 in 2019-2020 (October to January), 26 in 2020-2021 (December to January), 
22 in 2022 (January) and 23 in 2022-2023 (November to April), with a single sequence collected in most clear nights (except in one night where the sequence was repeated 
due to weather instabilities).  Four spectra were discarded in the first season as a result of an instrument problem and another 6 spectra were left out in the 
last season due to very low signal to noise ratios (SNRs), finally yielding a total of 79 validated Stokes $I$, $V$ and $N$ spectra of V347~Aur.  This 
series includes 14, 26, 22 and 17 such spectra for the 4 seasons, respectively spanning intervals of 42, 98, 23 and 160~d, and altogether covering a 
temporal window of 1258~d.  The full log of our observations is provided in Table~\ref{tab:log} of Appendix~\ref{sec:appA}.   

Our SPIRou spectra were all processed with \texttt{Libre ESpRIT}, the nominal reduction pipeline of ESPaDOnS at CFHT, optimized for spectropolarimetry 
and adapted for SPIRou \citep{Donati20}.  Least-Squares Deconvolution \citep[LSD,][]{Donati97b} was then applied to all reduced spectra, using a line mask 
constructed from the VALD-3 database \citep{Ryabchikova15} for $\teff=3500$~K and $\logg=3.5$ adapted to V347~Aur (see Sec~\ref{sec:par}).  
Atomic lines deeper than 10 per cent of the continuum level were selected, for a total of $\simeq$1500 lines with average wavelength and Land\'e factor of 
1750~nm and 1.2 respectively.  The noise levels $\sigma_V$ in the resulting Stokes $V$ LSD profiles range from 1.9 to 4.5 (median 2.5, in units of $10^{-4} I_c$ 
where $I_c$ denotes the continuum intensity).  We also applied LSD with a mask containing the CO lines of the CO bandhead (at 2.3~\mic) only, to obtain veiling 
estimates in the $K$ band ($r_K$), in addition to those for the whole spectrum derived from LSD profiles of atomic lines ($r_{JH}$, see Sec.~\ref{sec:eml} and 
Table~\ref{tab:log}).  Phases and rotation cycles were derived assuming a rotation period of $\Prot=4.12$~d (see Sec.~\ref{sec:bl} and Table~\ref{tab:par}) 
and counting from an arbitrary starting Barycentric Julian Date (BJD) of 2458688.0 (i.e., prior to our first SPIRou observation).  

\begin{figure}
\centerline{\includegraphics[scale=0.6,bb=20 20 380 390]{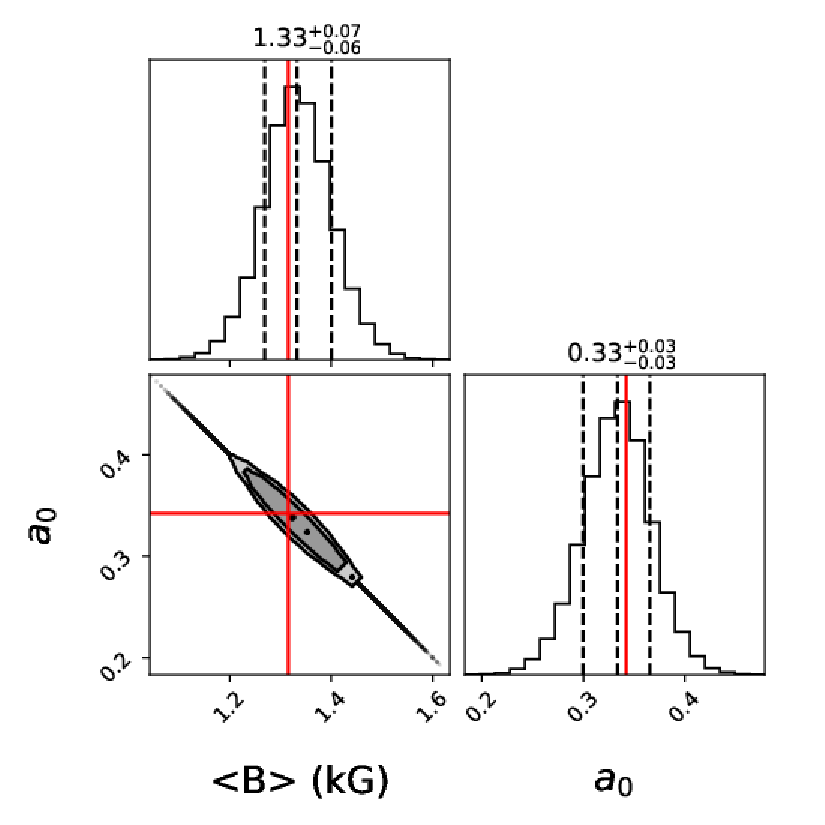}}
\caption[]{Magnetic parameters of V347~Aur, derived by fitting our median SPIRou spectrum using the atmospheric modeling approach of \citet{Cristofari23},
which incorporates magnetic fields as well as a MCMC process to determine optimal parameters and their error bars.  We find that V347~Aur hosts an average 
small-scale magnetic field of <$B$>~$=1.33\pm0.07$~kG, with non-magnetic regions covering a relative area of $a_0=33\pm3$~per cent of the visible stellar disc.  }
\label{fig:bmag}
\end{figure}

We also constructed a template spectrum of V347~Aur by computing the median of all 79 SPIRou spectra in the stellar rest frame \citep[with a method similar to that 
described in][]{Cristofari22a}, a process that allows not only to reduce photon noise and boost SNR, but also to get rid of OH airglow emission lines and of most 
telluric features (except the strongest ones).  We then carried out a spectrum characterization study using the ZeeTurbo code developed for SPIRou observations of 
M dwarfs \citep{Cristofari23,Cristofari23b} and adapted to the specific case of PMS stars and to V347~Aur in particular.  The parameter space is explored 
through a Monte-Carlo Markov Chain (MCMC) process, yielding posterior distributions and error bars for all parameters.  Assuming solar metallicity, we find 
atmospheric parameters equal to $\teff=3340\pm50$~K and $\logg=3.30\pm0.10$, consistent with those derived by \citet{Flores19}, and average veiling values 
$r_{JH}\simeq0.4$ and $r_K\simeq1.3$, in agreement with our measurements from LSD profiles (see Sec.~\ref{sec:eml}).  {\emr Simultaneously with the atmospheric 
parameters}, ZeeTurbo allows us to obtain an estimate of the average small-scale field at the surface of the star, found to be equal to $1.33\pm0.07$~kG (see 
Fig.~\ref{fig:bmag}) and consistent with \citet{Flores19}.  Given the limited SNR of our data, we carried out this analysis on the template spectrum of V347~Aur 
rather than on the individual spectra \citep[as in, e.g.,][]{Donati23}.  {\emr A portion of the observed and modeled spectra are shown as an example in 
Fig.~\ref{fig:spc}, along with the corresponding posterior distributions of the main parameters in Fig.~\ref{fig:corp}, of Appendix~\ref{sec:appB}}.  

\section{Spectropolarimetry of V347~Aur}
\label{sec:bl}

\begin{figure*}
\centerline{\includegraphics[scale=0.6,angle=-90]{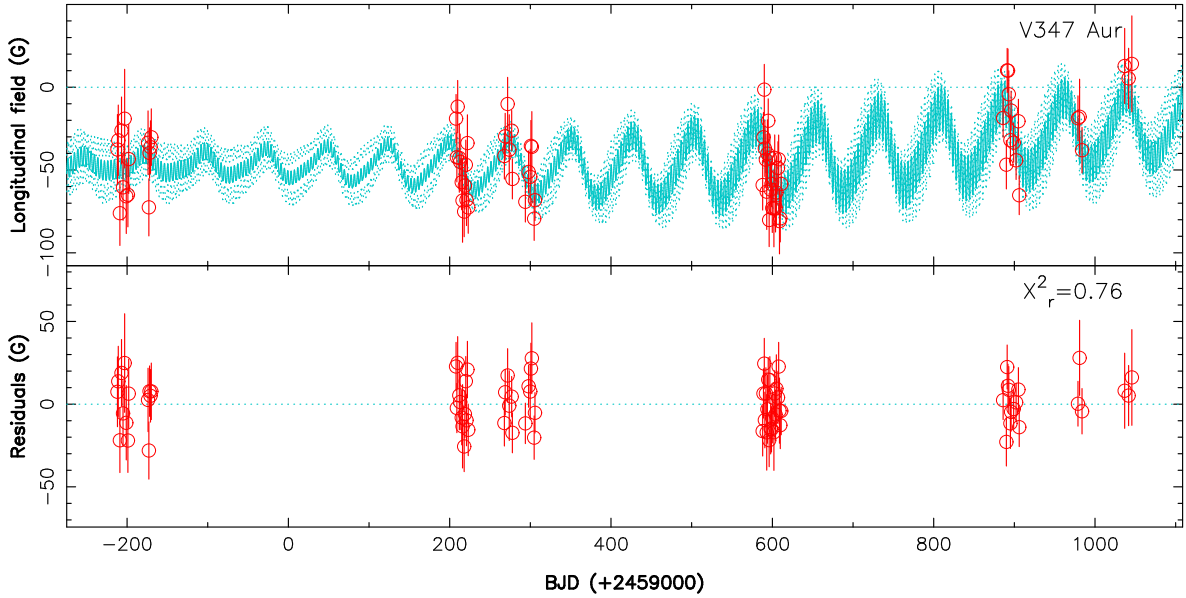}}
\caption[]{Longitudinal magnetic field \Bl\ of V347~Aur (red open circles) as measured with SPIRou (top plot), and QP GPR fit to the data (cyan full line) with 
corresponding 68~per cent confidence intervals (cyan dotted lines).  The residuals are shown in the bottom plot.   The rms of the residuals is 13.6~G ($\chisqr=0.76$), 
slightly smaller than our median error bar (15.3~G).  The \chisqr\ with respect to the $\Bl=0$~G line is equal to 9.7, indicating a clear magnetic detection. 
QP fluctuations of \Bl\ on a 76.5~d period are also detected, along with a weaker rotational modulation on a 4.2~d period.  }  
\label{fig:gpb}
\end{figure*}

We then examined our Stokes $I$ and $V$ LSD profiles and detected the presence of clear, though weak, Zeeman signatures in the spectral lines of V347~Aur.  To quantify 
our results, we derived 
the longitudinal field \Bl, i.e., the line-of-sight projected component of the vector field at the surface of the star averaged over the visible stellar hemisphere, 
for each pair of Stokes $V$ and $I$ LSD profiles, following \citet{Donati97b}.  We achieved this by computing the first moment of the Stokes $V$ profile and its 
error bar, which we normalised with the equivalent width of the Stokes $I$ LSD profiles estimated through a standard Gaussian fit.  In the case of V347~Aur, 
Stokes $V$ LSD signatures were integrated over a window of $\pm24$~\kms\ in the stellar rest frame, an interval adjusted to the width of stellar lines 
and corresponding to about twice the \vsini\ on both sides of the line center (the exact integration width having little impact on the result).  We find that 
the longitudinal field of V347~Aur is unambiguously detected, with moderate \Bl\ values ranging from --81 to +14~G (median --43~G) and error bars of 12 to 30~G 
(median 15~G), yielding a reduced chi-square \chisqr\ (with respect to the $\Bl=0$~G line) equal to $\chisqr=9.7$.  
We proceeded in the same way with the polarization check $N$ to verify that the associated pseudo longitudinal field is consistent with 0 within the error bars 
and find that $\chisqr=1.02$, indicating that no spurious pollution is observed in $N$ and that our analytical error bars are consistent with the observed 
measurement dispersion within a few per cent.  We note in particular that the longitudinal field of V347~Aur is smaller than those of most prototypical classical or 
weak-line T~Tauri stars (e.g., LkCa~4, CI~Tau, TW~Hya, GM~Aur) or more evolved PMS stars (e.g., AU~Mic) whose fields often significantly exceed 100~G in the nIR 
\citep[e.g.,][]{Finociety23,Donati23,Donati24,Donati24b,Zaire24} whereas that of V347~Aur stagnates at about 50~G or less, suggesting that the 
parent large-scale field is also smaller.  This is consistent with our result and that of \citet{Flores19} on the small-scale field of V347~Tau, also found to be 
weaker by a factor of $\simeq$2 than those of more evolved counterparts.  

In a second step, we studied whether temporal variability, including rotational modulation, is detected in our sequence of \Bl\ measurements.  {\emr The rms dispersion 
of \Bl\ values about their weighted average, equal to 23.5~G ($\chisqr=2.29$), is inconsistent with our error bars, indicating clear temporal variability.} 
A straightforward periodogram indicates that our \Bl\ series is dominated by a periodic signal at a period of about 76~d {\emr with a false-alarm probability of 0.02~per cent}.  
Given the main stellar parameters of V347~Aur, 
in particular \vsini\ and \rstar, yielding $\Prot<9.3$~d, this long period signal cannot be attributed to rotational modulation, and rather reflects an intrinsic periodic 
variability of the large-scale field, likely akin the well-known solar magnetic cycle but on a much shorter period.  To better characterize its properties, we 
investigated the temporal behaviour of our \Bl\ values, arranged in a vector denoted $\bf y$, using quasi-periodic (QP) Gaussian-Process Regression (GPR), with a 
covariance function $c(t,t')$ of type: 
\begin{eqnarray}
c(t,t') = \theta_1^2 \exp \left( -\frac{(t-t')^2}{2 \theta_3^2} -\frac{\sin^2 \left( \frac{\pi (t-t')}{\theta_2} \right)}{2 \theta_4^2} \right) 
\label{eq:covar}
\end{eqnarray}
where $\theta_1$ is the amplitude (in G) of the Gaussian Process (GP), $\theta_2$ its recurrence period, $\theta_3$ the evolution timescale 
(in d) on which the shape of the \Bl\ modulation changes, and $\theta_4$ a smoothing parameter describing the amount of harmonic complexity needed to describe the data \citep{Rajpaul15}.  
We then select the QP GPR fit that features the highest likelihood $\mathcal{L}$, defined by: 
\begin{eqnarray}
2 \log \mathcal{L} = -n \log(2\pi) - \log|{\bf C+\Sigma+S}| - {\bf y^T} ({\bf C+\Sigma+S})^{-1} {\bf y}
\label{eq:llik}
\end{eqnarray}
where $\bf C$ is the covariance matrix for our 79 epochs, $\bf \Sigma$ the diagonal variance matrix associated with $\bf y$, and ${\bf S}=\theta_5^2 {\bf J}$ ($\bf J$ 
being the identity matrix) the contribution from an additional white noise source that we introduce as a fifth hyper-parameter $\theta_5$ (in case \Bl\ is affected 
by intrinsic variability not included in our analytical error bars).  The hyper-parameter domain is then explored with a MCMC process, 
and the marginal logarithmic likelihood $\log \mathcal{L}_M$ of a given solution is computed using the approach of \citet{Chib01} as in, e.g., \citet{Haywood14}.  
This is achieved with the same MCMC and GPR modeling tools of our previous studies \citep[e.g.,][]{Donati23,Donati23b,Donati24,Donati24b}.  

The fit we obtain, associated with $\chisqr=1.13$ and $\log \mathcal{L}_M=-347.6$, {\emr yields a much smaller rms (16.5~G) than that about the weighted average of 
\Bl\ values (rms 23.5~G, $\chisqr=2.29$)}, and shows a definite improvement over a fit where only the longest term variation is adjusted \citep[i.e., with $\theta_2$ 
fixed to 1500~d, as in][yielding $\chisqr=1.69$, {\emr an rms of 20.2~G} and $\log \mathcal{L}_M=-356.6$]{Donati23b}.  The associated logarithmic Bayes 
factor $\log {\rm BF} = \Delta \log \mathcal{L}_M = 9.0$ thus demonstrates that the periodic fluctuation of \Bl\ at a period of 76.5~d is definitely real.  
The periodogram of the residuals now shows a peak at 4.2~d with a false-alarm probability of 0.6~per cent, consistent with what we expect for the rotation 
period of V347~Aur.  To include this second component into our analysis, we added another QP GP to our model, i.e., a new similar QP term to the covariance 
function of Eq.~\ref{eq:covar} defined again by 4 hyper parameters (called $\theta_6$ to $\theta_9$, and describing respectively the amplitude, the recurrence period, 
the evolution timescale and the smoothing parameter of the second GP).  

With this dual GP model, we further improve the match to our \Bl\ data (see Fig.~\ref{fig:gpb}), reaching now $\chisqr=0.76$ and $\log \mathcal{L}_M=-343.6$, i.e., 
$\log {\rm BF}=4.0$ compared to the single GP model including the 76.5~d component only.  Although short of the usual $\log {\rm BF}$ threshold (of 5) to firmly 
validate a detection, we nonetheless conclude that the 4.2~d component we see in the data, showing up as the only significant peak in the periodogram in the range 2-9~d, 
is most likely the rotation period of V347~Tau.  This preliminary conclusion is confirmed with our RV analysis of V347~Aur (see Sec.~\ref{sec:rvs}).  The final values 
and error bars derived from our GPR fit are listed in Table~\ref{tab:gpr}.  Note that both evolution timescales and smoothing parameters, loosely constrained by the 
\Bl\ data, were fixed to their most likely value, with little impact on the final result.  

We find in particular that both the 76.5~d cyclic fluctuation and the 4.2~d rotational modulation of \Bl\ increase with time over the duration of our observations (see  Fig.~\ref{fig:gpb}).  
Moreover, our result already suggests that the large-scale field of V347~Aur, in addition to being weaker than that of most prototypical classical T~Tauri and older PMS 
stars, is also more axisymmetric, with longitudinal fields keeping the same sign with rotation phase and exhibiting only marginal rotational modulation (of 
amplitude about twice smaller than that of the main 76.5~d fluctuation, see Table~\ref{tab:gpr}).  
We finally note that the 76.5-d fluctuation timescale of \Bl\ is consistent with the half recurrence period of the accretion outbursts and photometric pulses reported 
for V347~Aur \citep{Dahm20}, most likely not a coincidence.  We also investigated whether the outburst recurrence period itself was in fact the true period of the 
medium term fluctuations of \Bl, but found this hypothesis less probable (with $\log {\rm BF}=-2.5$).  We come back on this point in Sec.~\ref{sec:dis}.  

\begin{table} 
\caption[]{Results of our MCMC modeling of the \Bl\ curve of V347~Aur with GPR involving 2 QP GPs, the first one for the 76.5~d \Bl\ fluctuation and the second one for 
the rotational modulation.  For each hyper-parameter, we list the fitted value, the corresponding 
error bar and the assumed prior.  The knee of the modified Jeffreys prior is set to $\sigma_{B}$, i.e., the median error bar of our \Bl\ measurements (i.e., 15.3~G). 
{\emr In the final two rows, we also successively quote the \chisqr\ and rms of the \Bl\ data about the weighted average, once fitted by a single GP (hyper parameters 
$\theta_1$ to $\theta_5$) and by our dual GP model.} }  
\scalebox{0.95}{\hspace{-2mm}
\begin{tabular}{ccccc}
\hline
Parameter   & Name & Value & Prior   \\
\hline 
Amplitude (G)        & $\theta_1$  & $22^{+8}_{-6}$  & mod Jeffreys ($\sigma_{B}$) \\
Rec.\ period (d)     & $\theta_2$  & $76.5\pm1.2$   & Gaussian (76.5, 5.0) \\
Evol.\ timescale (d) & $\theta_3$  & 500             & fixed \\
Smoothing            & $\theta_4$  & $0.75$          & fixed \\
\hline 
White noise (G)      & $\theta_5$  & $6\pm3$         & mod Jeffreys ($\sigma_{B}$) \\
\hline 
Amplitude (G)        & $\theta_6$  & $12^{+6}_{-4}$  & mod Jeffreys ($\sigma_{B}$) \\
Rec.\ period / \Prot\ (d)     & $\theta_7$  & $4.19\pm0.16$   & Gaussian (4.2, 0.5) \\
Evol.\ timescale (d) & $\theta_8$  & 300             & fixed \\
Smoothing            & $\theta_9$  & $0.75$          & fixed \\
\hline
        \multicolumn{4}{c}{\emr Goodness of fit} \\ 
                     & \emr vs average  & \emr single GP  & \emr dual GP    \\ 
\chisqr              & \emr 2.29        & \emr 1.13            & 0.76       \\ 
rms (G)              & \emr 23.5        & \emr 16.5            & 13.6       \\ 
\hline 
\end{tabular}}      
\label{tab:gpr}      
\end{table}

\section{RV analysis and orbital parameters of V347~Aur}
\label{sec:rvs}

In the next step, we focussed on our Stokes $I$ LSD profiles of V347~Aur, from which we extracted RVs and corresponding error bars for our 79 data points.  We achieved this by 
describing each individual LSD profile as a simple first order Taylor expansion constructed from the median of all LSD profiles.  This approach is similar to that used in the 
'line-by-line' technique \citep{Artigau22}, but applied to our LSD profiles (corrected from veiling, see Sec.~\ref{sec:zdi}) rather than to individual spectral lines.  In 
practice, we start by fitting the median LSD profile with a Gaussian, then adjust each differential LSD profile with respect to the median with a scaled version of the first 
derivative (with respect to velocity) of the Gaussian fit to the median LSD profile.  The scaling factors directly yield the RV shifts of our observations with respect to the 
median LSD profile, with the corresponding error bars on the RV shifts being computed from the SNRs, the full widths at half maximum and the equivalent widths (EWs) of the 
individual LSD profiles.  We find this approach to be slightly more precise and stable than a straightforward Gaussian fit to our LSD profiles, thanks in particular to the 
cleaner continuum levels one gets by dividing each LSD profile by the median.  The derived RVs, ranging from 6.8 to 9.5~\kms\ (median 8.6~\kms), and associated error bars 
(median 0.2~\kms) are listed in Table~\ref{tab:log}.  We note in particular that V347~Aur is indeed RV variable, with a peak-to-peak variation exceeding 2~\kms\ (i.e., 
10$\times$ our median RV error bar), thereby confirming the suspicion of \citet{Dahm20} that it is a spectroscopic binary.  

\begin{figure*}
\centerline{\includegraphics[scale=0.6,angle=-90]{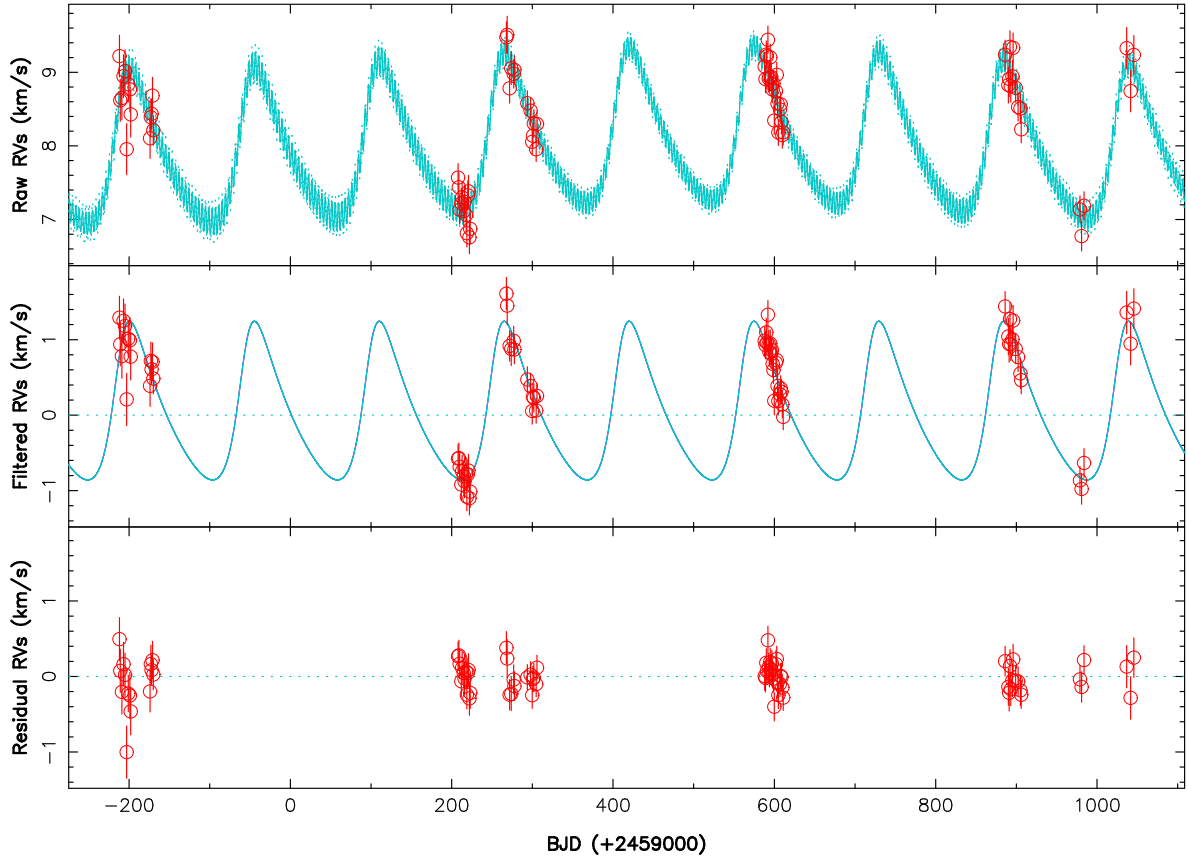}} 
\caption[]{Raw (top), activity filtered (middle) and residual (bottom) RVs of V347~Aur (red open circles).  The top plot shows the MCMC fit to the RV data, including 
a QP GPR modeling of the activity and the orbital motion of the primary star (cyan full line, with cyan dotted lines illustrating the 68~per cent confidence intervals), 
whereas the middle plot shows the orbital motion only once activity is filtered out.  The rms of the RV residuals is 0.19~\kms. } 
\label{fig:rvr}
\end{figure*}

\begin{figure}
\includegraphics[scale=0.48,angle=-90]{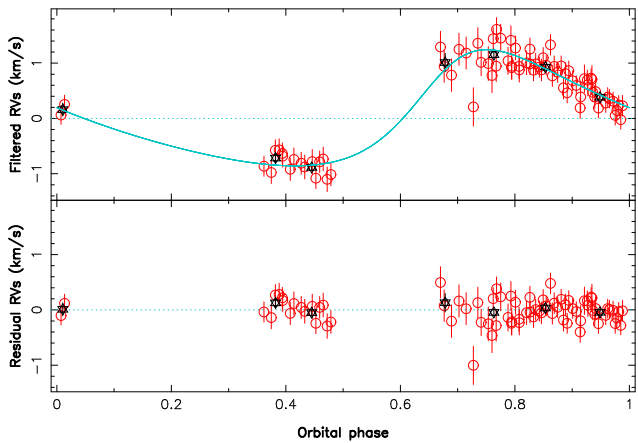}
\caption[]{Filtered (top plot) and residual (bottom plot) RVs of V347~Aur phase-folded on the 154.6~d period.  
The red open circles are the individual RV points with the respective error bars, whereas the black stars are average RVs over 0.1 phase bins.
As in Fig.~\ref{fig:rvr}, the dispersion of RV residuals is 0.19~\kms. }
\label{fig:rvf}
\end{figure}

\begin{table}
\caption[]{MCMC results of our RV analysis of V347~Aur. We list the recovered GP and orbital parameters with their error bars (with T$_{\rm c}$ and T$_{\rm p}$ denoting 
the times of inferior conjunction and periastron passage respectively, $a$ the semi-major axis of the orbit, and $a_{\rm p}$ and $a_{\rm a}$ the distances between both 
components at periastron and apoastron respectively), as well as the corresponding priors whenever relevant.  The last 2 rows give the \chisqr\ and the rms of the best fit 
to our RV data. } 
\scalebox{0.95}{\hspace{-2mm}
\begin{tabular}{cccccccc}
\hline
Parameter          & Value                &   Prior \\
\hline
$\theta_1$ (\kms)  & $0.22^{+0.07}_{-0.06}$ & mod Jeffreys ($\sigma_{\rm RV}$) \\
$\theta_2$ (d)     & $4.12\pm0.03$          & Gaussian (4.12, 0.2) \\
$\theta_3$ (d)     & 300                    &  \\
$\theta_4$         & 0.85                   &  \\
$\theta_5$ (\kms)  & $0.08\pm0.06$          & mod Jeffreys ($\sigma_{\rm RV}$) \\
\hline
$K$ (\kms)         & $1.05\pm0.06$          & mod Jeffreys ($\sigma_{\rm RV}$) \\
\Porb\ (d)         & $154.6\pm0.7$          & Gaussian (155, 5) \\
T$_{\rm c}$ (2459000+) & $457.5\pm3.0$            & Gaussian (458, 10) \\
T$_{\rm p}$ (2459000+) & $408.0\pm4.2$            & from orbital elements \\
$\sqrt e\cos\omega$& $0.33\pm0.08$          & Gaussian (0, 1)       \\ 
$\sqrt e\sin\omega$& $-0.42\pm0.12$         & Gaussian (0, 1)       \\ 
$e$                & $0.28\pm0.12$          & from $\sqrt e\cos\omega$ and $\sqrt e\sin\omega$& \\ 
$\omega$ (rad)     & $-0.91\pm0.18$         & from $\sqrt e\cos\omega$ and $\sqrt e\sin\omega$& \\ 
$M_b \sin i$ (\mjup) & $12.7\pm0.7$         & from $K$, \Porb\ and \mstar \\
$M_b$ (\mjup)      & $29.0\pm1.6$           & from $M_b \sin i$ and $i$ \\
$a$ (au)           & $0.39\pm0.02$          & from \Porb, \mstar\ and $e$ \\ 
$a_{\rm p}$ (au)   & $0.28\pm0.05$          & from $a$ and $e$ \\ 
$a_{\rm a}$ (au)   & $0.50\pm0.05$          & from $a$ and $e$ \\ 
\hline
\chisqr            & 0.93   & \\
rms (\kms)         & 0.19   & \\
\hline
\end{tabular}}
\label{tab:pla}
\end{table}

We modeled our RV data with a Keplerian function describing the orbital motion of the primary star, along with a (single) QP GP to account for the activity jitter.  
The orbital motion is very clearly detected, with a semi-amplitude $K=1.05\pm0.06$~\kms\ measured at a 17.5$\sigma$ confidence level and an orbital period 
$\Porb=154.6\pm0.7$~d fully consistent with the estimate of \citet[][also 154.6~d with no quoted error bar]{Dahm20}.  Although our phase coverage of the orbit is still 
incomplete, we can safely claim that the orbital motion is eccentric, with an ellipticity of $e=0.28\pm0.12$ and a difference in $\log {\rm BF}$ reaching 9.4 in favour 
of the elliptic case compared to the circular one.  

The activity jitter is also clearly detected despite its semi-amplitude (0.22~\kms, see Table~\ref{tab:pla}) being only slightly larger than our median error bar, the 
$\log {\rm BF}$ in favour of the model including a GP reaching now 8.5 compared to that featuring no GP.  Besides, this model also yields a more accurate value of 
\Prot\ ($\Prot=4.12\pm0.03$~d) than that derived from our \Bl\ data (see Sec.~\ref{sec:bl}), and further confirms that both \Bl\ and RV exhibit consistent rotational 
modulation.  The results of our analysis are depicted in Fig.~\ref{fig:rvr} whereas the derived parameters are listed in Table~\ref{tab:pla}.  We also show in 
Fig.~\ref{fig:rvf} our RV data and inferred orbital RV curve phase-folded on the derived orbital period of 154.6~d.  

The inferred minimum mass for the companion is equal to $M_b \sin i=12.7\pm0.7$~\mjup, at the limit between the planet and brown dwarf domain.  If we further
assume that the companion orbits within the equatorial plane of the star (presumably coinciding with the plane of the accretion disc) 
we end up with a companion mass of $M_b=29.0\pm1.6$~\mjup\ and a mass ratio $q$ of the system equal to $q=0.009\pm0.003$.  This makes the companion a bona fide brown 
dwarf, orbiting V347~Aur with a semi-major axis $a=0.39\pm0.02$~au, and with periastron and apoastron distances equal to $a_{\rm p}=0.28\pm0.05$~au 
and $a_{\rm a}=0.50\pm0.05$~au respectively.

\section{Zeeman-Doppler Imaging of V347~Aur}
\label{sec:zdi}

We then analysed the Stokes $I$ and $V$ LSD signatures of V347~Aur using ZDI, to investigate the large-scale magnetic topology of V347~Aur and its evolution with time 
over the duration of our observations.  As ZDI usually works with the assumption that the observed variability mostly relates to rotational modulation, and given 
that the longitudinal field of V347~Aur was found to also fluctuate with a period longer than \Prot\ ($76.5\pm1.2$~d, see Sec.~\ref{sec:bl}), we chose to decompose 
our full data set into 7 subsets, respectively including observations from 2019 Oct-Dec (14 spectra), 2020 Dec-2021 Jan (13 spectra), 2021 Feb-Apr (13 spectra), 
2022~Jan (22 spectra), 2022~Nov (11 spectra), 2023~Feb (3 spectra) and 2023 Apr (3 spectra), each lasting 5--10 rotation cycles (except for the 2 final ones lasting only 
1--2 rotation cycles).  Obviously, the last 2 subsets do not include enough observations to yield fully reliable ZDI images, but we nonetheless chose to analyse them as such, 
keeping in mind the limitations associated with the sparse coverage.  
These separate subsets also correspond to different phases of the orbital cycle of V347~Aur (see Sec.~\ref{sec:rvs}), thereby allowing us to 
study how the large-scale field evolves on the previously mentioned timescales.  

The ZDI code used here, already applied several times to SPIRou Stokes $V$ data sets of T~Tauri stars \citep[e.g.,][]{Finociety21,Finociety23, Donati24,Donati24b,Zaire24}, 
allows one to reconstruct the topology of the large-scale field at the surface of a star from phase-resolved sets of Stokes $V$ LSD profiles.  
This is achieved with an iterative process that progressively adds information at the surface of the star, starting from a small magnetic seed and exploring the 
parameter space in order to minimize the discrepancy between observed and synthetic data \citep[e.g.,][]{Brown91, Donati97c, Donati06b}.  As this problem is ill posed, 
regularization is required to ensure a unique solution.  ZDI uses the principles of maximum entropy image reconstruction \citep{Skilling84} to reach a given agreement 
with the data, usually $\chisqr\simeq1$, with the minimum amount of information in the reconstructed image.  

Technically speaking, the surface of the star is modeled as a grid of 5000 cells.  We then compute the spectral contributions of each grid cell, using Unno-Rachkovsky's 
analytical solution of the polarized radiative transfer equation in a plane-parallel Milne Eddington atmosphere \citep{Landi04}, and assuming a Land\'e factor,  
average wavelength and Doppler width of 1.2, 1750~nm and 3~\kms\ for the local profile (as in our previous studies).  We then sum up the contributions of all grid 
cells, taking into account the star and cell characteristics and assuming the star rotates as a solid body, to derive the synthetic profiles at the observed phases 
of the rotation cycle.  The magnetic field at the surface of the star is expressed as a spherical harmonic expansion, using the formalism of \citet[][see also 
\citealt{Lehmann22,Finociety22,Donati23}]{Donati06b} in 
which the poloidal and toroidal components of the vector field depend on 3 sets of complex spherical harmonic coefficients, $\alpha_{\ell,m}$ and $\beta_{\ell,m}$ for the poloidal 
component, and $\gamma_{\ell,m}$ for the toroidal component, where $\ell$ and $m$ note the degree and order of the corresponding spherical harmonic term in the expansion.  
Given the moderate \vsini\ of V347~Aur and the weak detected magnetic signatures (see Sec.~\ref{sec:bl}), the spherical harmonic expansion is limited to terms up to $\ell=5$.  

\begin{figure*}
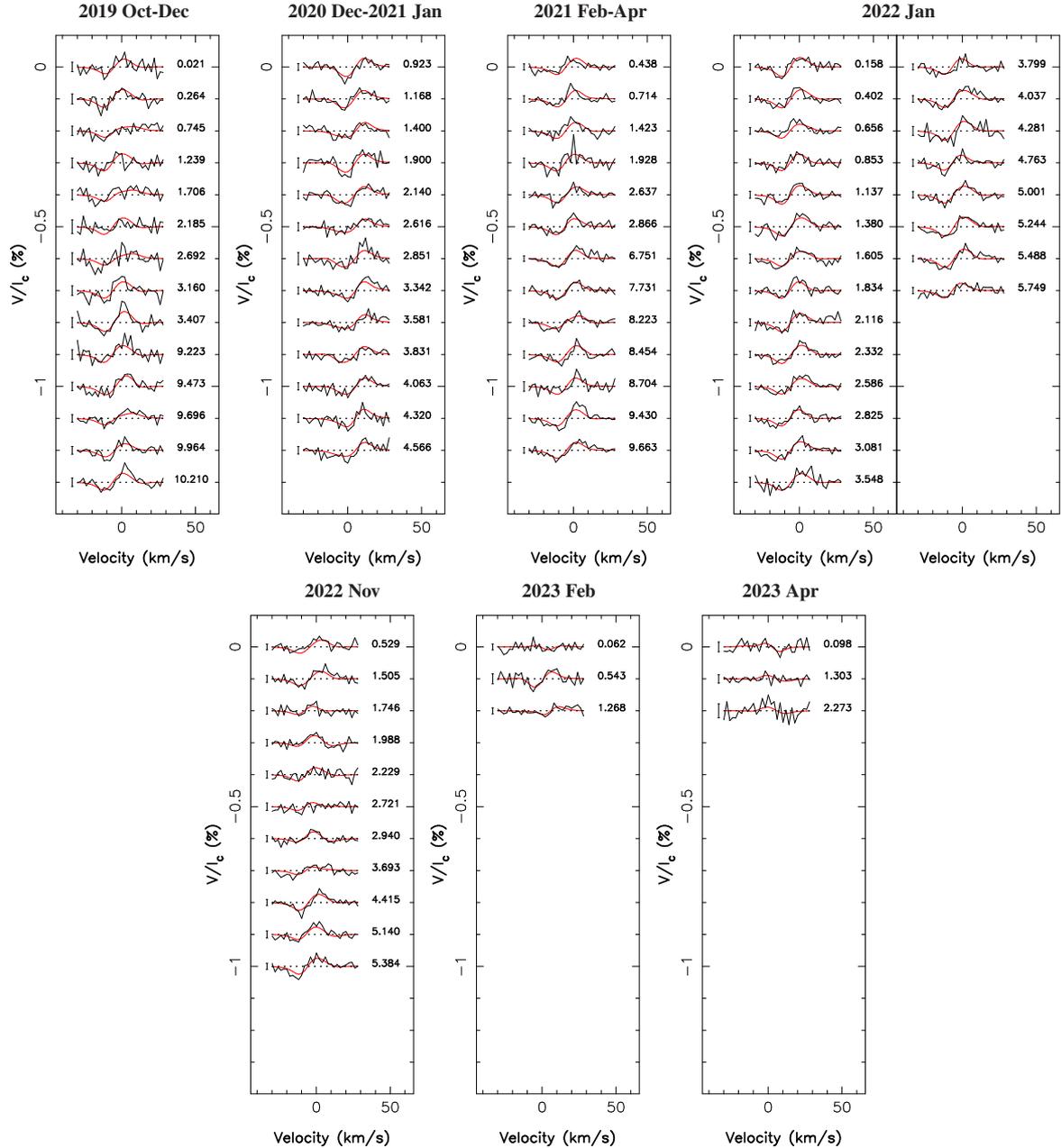

\flushleft{\bf \hspace{2.2cm}2019 Oct-Dec \hspace{1.3cm} 2020 Dec-2021 Jan  \hspace{1.2cm} 2021 Feb-Apr \hspace{3cm} 2022 Jan\vspace{-1mm}} 
\centerline{\includegraphics[scale=0.45,angle=-90]{fig/v347aur-fit19.ps}\hspace{2mm}\includegraphics[scale=0.45,angle=-90]{fig/v347aur-fit20a.ps}\hspace{2mm}\includegraphics[scale=0.45,angle=-90]{fig/v347aur-fit20b.ps}\hspace{2mm}\includegraphics[scale=0.45,angle=-90]{fig/v347aur-fit21.ps}\vspace{7mm}} 
\flushleft{\bf \hspace{5.5cm}2022 Nov\hspace{2cm} 2023 Feb \hspace{2cm} 2023 Apr\vspace{-1mm}} 
\centerline{\includegraphics[scale=0.45,angle=-90]{fig/v347aur-fit22a.ps}\hspace{2mm}\includegraphics[scale=0.45,angle=-90]{fig/v347aur-fit22b.ps}\hspace{2mm}\includegraphics[scale=0.45,angle=-90]{fig/v347aur-fit22c.ps}} 
\caption[]{{\emr Observed (thick black line) and modeled (thin red line) LSD Stokes $V$ profiles} of V347~Aur in the stellar rest frame, for the first 4 
subsets (top row, corresponding to 2019 Oct-Dec, 2020 Dec - 2021 Jan, 2021 Feb-Apr and 2022 Jan respectively) and the last 3 subsets (bottom row, corresponding to 
2022 Nov, 2023 Feb and 2023 Apr respectively).  Observed profiles, obtained by applying LSD to SPIRou spectra (using the atomic line mask outlined in Sec.~\ref{sec:obs}), were 
corrected for the orbital motion described in Sec.~\ref{sec:rvs} and the veiling outlined in Sec.~\ref{sec:eml}.   
Note how the Stokes $V$ profiles of each subset are, on average, either blue- or red-shifted with respect to the line centre, indicating the presence of a strong toroidal field 
(see text).  Rotation cycles (counting from 0, 101, 116, 194, 266, 289 and 303 for subsets \#1 to \#7 respectively, see Table~\ref{tab:log}) are indicated to the right of all 
LSD profiles, while $\pm$1$\sigma$ error bars are added to the left of each profile. }
\label{fig:fit}
\end{figure*}

\begin{figure*}
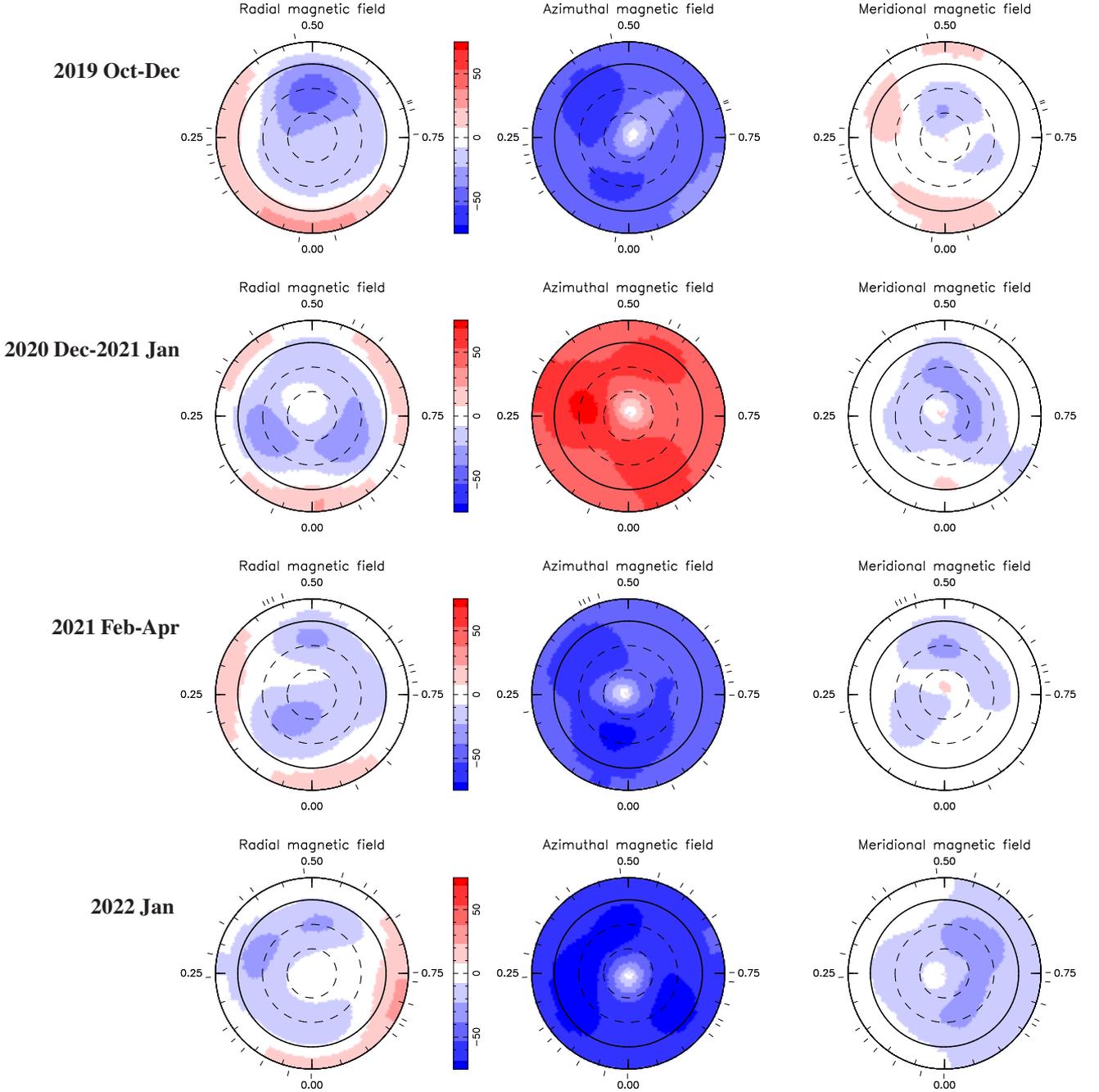

\flushright{\large\bf 2019 Oct-Dec\raisebox{0.3\totalheight}{\includegraphics[scale=0.4,angle=-90]{fig/v347aur-map19-3.ps}}\vspace{2mm}}
\flushright{\large\bf 2020 Dec-2021 Jan\raisebox{0.3\totalheight}{\includegraphics[scale=0.4,angle=-90]{fig/v347aur-map20a-3.ps}}\vspace{2mm}}
\flushright{\large\bf 2021 Feb-Apr\raisebox{0.3\totalheight}{\includegraphics[scale=0.4,angle=-90]{fig/v347aur-map20b-3.ps}}\vspace{2mm}}
\flushright{\large\bf 2022 Jan    \raisebox{0.3\totalheight}{\includegraphics[scale=0.4,angle=-90]{fig/v347aur-map21-3.ps}}}
\caption[]{Reconstructed maps of the large-scale field of V347~Aur (left, middle and right columns for the radial, azimuthal and meridional components in spherical 
coordinates, in G), for our first 4 data subsets (top to bottom rows) derived with ZDI from the Stokes $V$ LSD profiles of Fig.~\ref{fig:fit}.  
The maps are shown in a flattened polar projection down to latitude $-30$\degr, with the north pole at the centre and the equator depicted as a bold line.  Outer ticks 
indicate phases of observations.  Positive radial, azimuthal and meridional fields respectively point outwards, counterclockwise and polewards. } 
\label{fig:map1}
\end{figure*}

\begin{figure*}
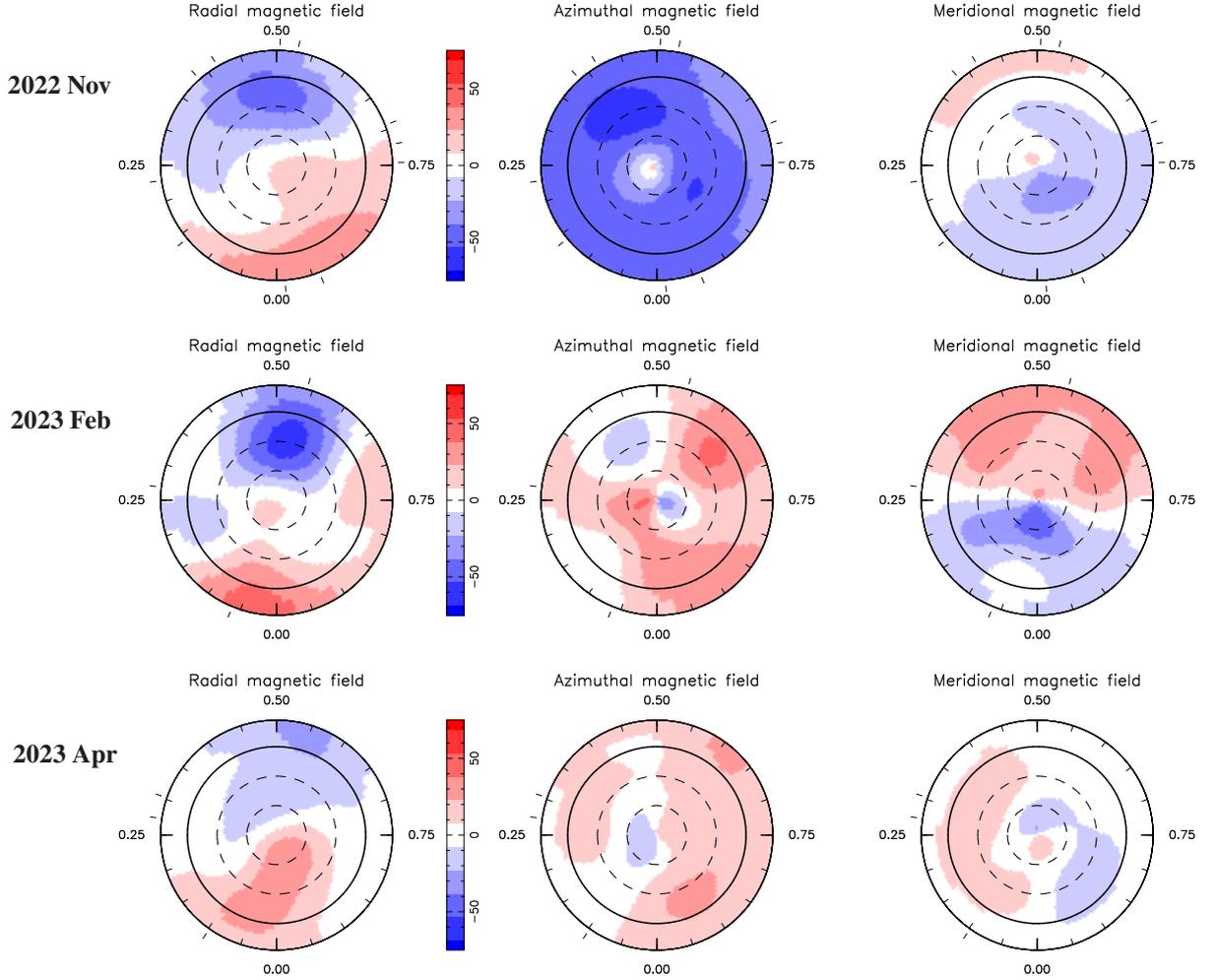

\flushright{\large\bf 2022 Nov \raisebox{0.3\totalheight}{\includegraphics[scale=0.4,angle=-90]{fig/v347aur-map22a-3.ps}}\vspace{2mm}}
\flushright{\large\bf 2023 Feb \raisebox{0.3\totalheight}{\includegraphics[scale=0.4,angle=-90]{fig/v347aur-map22b-3.ps}}\vspace{2mm}}
\flushright{\large\bf 2023 Apr\raisebox{0.3\totalheight}{\includegraphics[scale=0.4,angle=-90]{fig/v347aur-map22c-3.ps}}}
\caption[]{Same as Fig.~\ref{fig:map1} for our last 3 data subsets.  Note the limited phase coverage for the last 2 maps, especially the last one, rendering them 
much less reliable than the others.  } 
\label{fig:map2}
\end{figure*}

Before applying ZDI, we started by re-centring the Stokes $I$ and $V$ LSD profiles from the orbital motion outlined in Sec.~\ref{sec:rvs}.  Moreover, as spectra of 
V347~Aur are subject to veiling as a result of accretion (see Sec.~\ref{sec:eml}), we corrected all Stokes $I$ profiles by normalizing them to the same 
EW, and applied the derived scaling factors to the corresponding Stokes $V$ profiles as well.  Since intrinsic variability induced by variable accretion causes the 
normalized Stokes $I$ profiles to feature random distortions that ZDI is not able to easily reproduce, we chose not to attempt modeling them and 
only ensured that the median Stokes $I$ profile of each subset is consistent with the median synthetic profile of our model.  This modeling yielded in particular an estimate of 
\vsini, found to be equal to $\vsini=12.0\pm0.5$~\kms\ \citep[in agreement with][]{Flores19}.  We then ran ZDI on each subset of Stokes $V$ profiles to reconstruct the 
large-scale magnetic topology at each epoch.  The corresponding fits to the Stokes $V$ profiles, matched at a level of $\chisqr\simeq1$, are shown in Fig.~\ref{fig:fit} 
for our 7 subsets, whereas the reconstructed magnetic maps are depicted in Figs.~\ref{fig:map1} and \ref{fig:map2}.  

As already pointed out above (see Sec.~\ref{sec:bl}), we find that the reconstructed large-scale field of V347~Aur is quite weak for such a young active star, with an 
average strength of only about 50~G \citep[compared to 0.8--1.1~kG for CI~Tau and TW~Hya,][]{Donati24,Donati24b}.  Another surprise is that the derived large-scale field is 
mostly toroidal, with the poloidal component gathering only $\simeq$20~per cent of the reconstructed magnetic energy (except at the last 2 epochs, especially the last one, 
where phase coverage is limited and for which the maps are much less reliable).  This dominant toroidal field directly reflects that the average Zeeman signature over each subset 
is significantly shifted with respect to the line center \citep[see, e.g.,][]{Donati05,Lehmann22}.  This is especially obvious in the first 5 subsets where phase coverage is 
reasonably even, and for which the shift of the Stokes $V$ signatures with respect to the line center reaches about $+$5~\kms\ (in subset \#2) and $-$5~\kms\ (in all others, see 
Fig.~\ref{fig:fit}).   

\begin{table} 
\caption[]{Properties of the large-scale magnetic field of V347~Aur for our 7 data subsets.  
We list the average reconstructed large-scale field strength <$B_V$> (column 2), the polar strength of the dipole 
component \Bd\ (column 3), the tilt of the dipole field to the rotation axis and the phase towards which it is tilted (column 4) and the amount of magnetic energy reconstructed 
in the poloidal component of the field and in the axisymmetric modes of this component (column 5).  Error bars are typically equal to 5--10~per cent for field strengths and percentages, 
and 5--10\degr\ for field inclinations, for the first 5 subsets, and at least twice worse for the last 2 (for which ZDI is less reliable given the sparse phase covarage). } 
\begin{tabular}{ccccc}
\hline
Data subset       & <$B_V$> & \Bd     & tilt / phase    & poloidal / axisym \\ 
                  &  (G)    &  (G)    & (\degr)         & (per cent)        \\ 
\hline
2019 Oct-Dec      & 49 & 35 & 27 / 0.54 & 22 / 71 \\ 
2020 Dec-2021 Jan & 54 & 27 &  7 / 0.12 & 17 / 82 \\  
2021 Feb-Apr      & 52 & 23 & 13 / 0.71 & 11 / 76 \\  
2022 Jan          & 62 & 21 & 35 / 0.31 & 10 / 81 \\  
2022 Nov          & 48 & 31 & 77 / 0.44 & 20 / 23 \\  
2023 Feb          & 37 & 33 & 65 / 0.49 & 48 / 19 \\  
2023 Apr          & 21 & 23 & 74 / 0.01 & 70 /  8 \\  
\hline 
\end{tabular}
\label{tab:mag}
\end{table}

The dipole component of the large-scale field stores 60--85~per cent of the poloidal field energy, with a polar strength of only 
about 30~G.  The poloidal field is mainly axisymmetric for the first 4 epochs (up to 2022~Jan), but starts to significantly evolve in 2022~Nov, i.e., when the {\emr modulus of the} 
average longitudinal field progressively weakens (see Fig.~\ref{fig:gpb}) and the dipole field gets highly tilted to the rotation axis.  The dominant toroidal field is mostly 
axisymmetric at all times.  Moreover, it switches from negative to positive polarity between the first and second subsets (with the mean Zeeman signature shifting from the 
blue wing to the red wing of the line profile, see Fig.~\ref{fig:fit}) then switches sign again between the second and the third subsets.  It may do it one 
more time between the fifth and sixth subsets, although the sparse phase coverage of the last 2 epochs (and especially the last one covering less than a fourth of the 
rotation cycle) makes the latter conclusion far less firm.  The derived properties of the large-scale field for each subset are summarized in Table~\ref{tab:mag}.  

{\emr We note in particular that the toroidal field apparently switches sign as V347~Aur goes through periastron (between the second and third subsets), and again 
once it crosses apoastron (between the fifth and sixth subsets, and between the first and second, see Figs.~\ref{fig:rvr}, \ref{fig:map1} and \ref{fig:map2}).  
The expected sign switch between our penultimate and last subsets, coinciding again with V347~Aur passing through periastron, does not show up in our reconstructed images, } 
presumably because of the very poor phase coverage of our last subset, and possibly 
due to the large-scale field evolving on a longer timescale towards the end of our observations (with a longitudinal field getting closer to zero at this epoch, 
see Fig.~\ref{fig:gpb}).  

We also note that, in contrast with the toroidal field apparently switching sign with the orbital period, the poloidal field and its dominant dipole 
component do not exhibit similar polarity switches, at least not with the same period.  The increase in both the dipole tilt and the fractional poloidal field energy stored in 
non-axisymmetric components that we see towards the end of our observations rather suggests that the poloidal field only switches sign on a timescale longer than that of our 
campaign.  Similarly, we see no obvious change in the properties of the reconstructed large-scale field that correlates with the 76.5-d fluctuation of \Bl\  
(see Sec.~\ref{sec:bl}), apart from the one associated with the toroidal field polarity switches (apparently occurring on a twice longer period).    
We stress however that the temporal sampling of our 7 maps is likely 
too sparse to precisely document the evolution of the magnetic topology, and in particular of its poloidal component, with the 76.5~d period.  {\emr  We speculate that the 76.5-d 
fluctuation of \Bl, mostly sensitive to the poloidal field, actually reflects a forced periodic oscillation of this field component about a non-zero value, likely triggered by the 
orbital motion, and that a denser monitoring of V347~Aur is required for it to show up more convincingly on ZDI images. }

\section{Veiling and emission lines of V347~Aur}
\label{sec:eml}

To complete this analysis, we investigated accretion proxies of V347~Aur, through the veiling of photospheric lines and the profiles of well-known nIR emission lines, i.e., 
the 1083.3-nm \hei\ triplet, the 1282.16-nm \pab\ line and the 2166.12-nm \brg\ line.  

\begin{figure*}
\centerline{\includegraphics[scale=0.6,angle=-90]{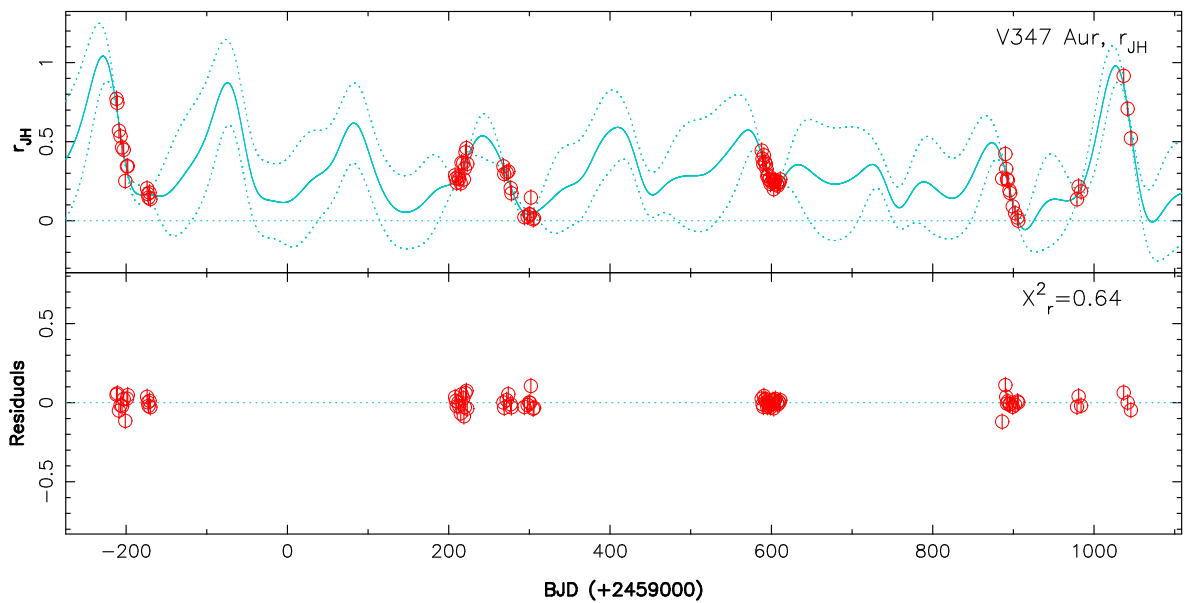}}
\caption[]{Veiling in $JH$ bands $r_{JH}$ of V347~Aur (top plot), as measured from LSD profiles of atomic lines (red open circles), and QP GPR fit to the data (cyan full line) 
with corresponding 68~per cent confidence intervals (cyan dotted lines).  The residuals are shown in the bottom plot. }  
\label{fig:vei}
\end{figure*}

To measure veiling, we compared LSD profiles of V347~Aur with those of TWA~10, a weak-line T~Tauri star of similar temperature (but different \logg) also monitored with 
SPIRou.  We derived the veiling estimates $r_{JH}$ from LSD profiles of atomic lines, and $r_K$ from LSD profiles of the CO bandhead lines (see Sec.~\ref{sec:obs}).  
In both cases, veiling estimates were derived by comparing the EWs of each LSD profile of 
V347~Aur with the corresponding median LSD profile of TWA~10, EWs being measured through standard Gaussian fits.  The resulting values of $r_{JH}$ and $r_K$ are listed in 
Table~\ref{tab:log} {\emr and shown in Fig.~\ref{fig:rjhk} in Appendix~\ref{sec:appC}}.  We find that $r_{JH}$ ranges from 0.0 to 0.92, with a median of 0.26, an average of 0.29 and an rms 
of 0.17, whereas $r_K$ ranges from 0.1 to 5.1, with a median of 0.73, an average of 1.00 and an rms of 0.81.  We also find that $r_K$ is well correlated ($R\simeq0.8$) with, 
and on average 2.9$\times$ stronger than, $r_{JH}$.  Our estimates are consistent with those of \citet{Sousa23}, derived from a subset of the same SPIRou spectra (with a different method).   

We analysed the temporal variations of $r_{JH}$ using the same approach as that used for \Bl, i.e., a QP GPR fit coupled to a MCMC process to derive the optimal hyper 
parameters describing the time series (see Sec.~\ref{sec:bl}).  The derived model, shown in Fig.~\ref{fig:vei}, exhibits peaks occurring with a period of $156\pm6$~d, 
consistent within error bars with the orbital period of V347~Aur derived from our RV analysis ($\Porb=154.6\pm0.7$~d, see Table~\ref{tab:pla}).  These peaks (e.g., the one at 
BJD~$\simeq2459405$) coincide with the periastron passage of V347~Aur as derived from our RV analysis (predicting $T_p=2459408.0\pm4.2$, see Table~\ref{tab:pla}).  
We note that the veiling maxima feature unequal strengths (by up to a factor of 2 in our case), qualitatively similar to what was reported for photometric 
maxima and interpreted as irregular pulsed-accretion events triggered by the orbital motion and occurring close to periastron passage \citep{Dahm20}.  Apparently, the 
two strongest such events in our data set, which at least partly sampled 5 of these maxima, occurred near the beginning and end of our observations.  We also note that 
$r_{JH}$ can go down to 0 between two accretion peaks.    

\begin{figure*}
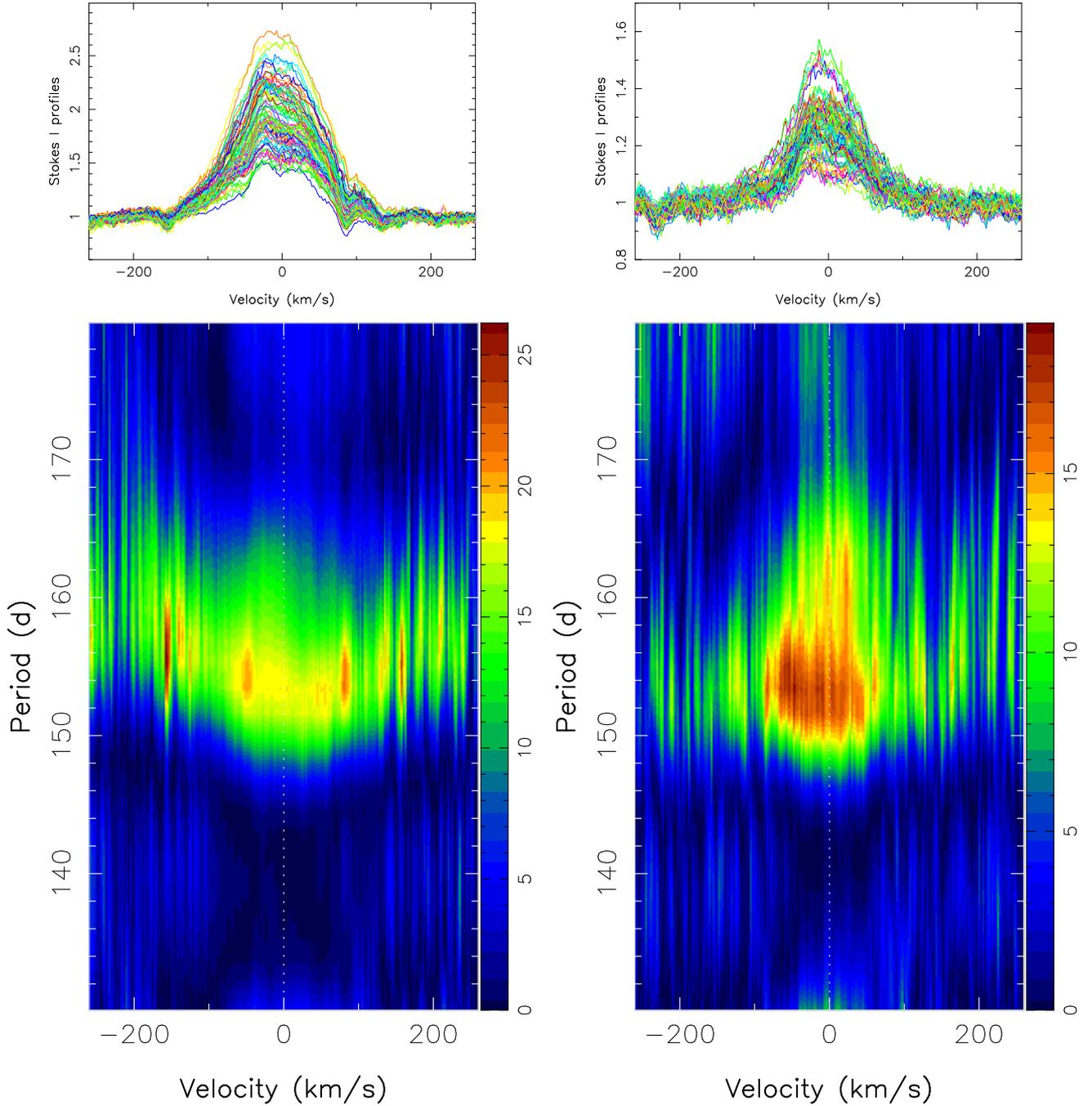

\centerline{\hspace{-2mm}\includegraphics[scale=0.3,angle=-90]{fig/v347aur-pab.ps}\hspace{19mm}\includegraphics[scale=0.3,angle=-90]{fig/v347aur-brg.ps}\vspace{2mm}}
\centerline{\includegraphics[scale=0.55,angle=-90]{fig/v347aur-pab-p2d.ps}\hspace{3mm}\includegraphics[scale=0.55,angle=-90]{fig/v347aur-brg-p2d.ps}}
\caption[]{Stacked Stokes $I$ profiles and 2D periodograms of the 1282.16-nm \pab\ line (left) and the 2166.12-nm \brg\ line (right) of V347~Aur, in the stellar 
rest frame, for our complete data set.  The color scale depicts the logarithmic power of the periodogram. } 
\label{fig:eml1}
\end{figure*}

\begin{figure*}
\centerline{\includegraphics[scale=0.7,angle=-90]{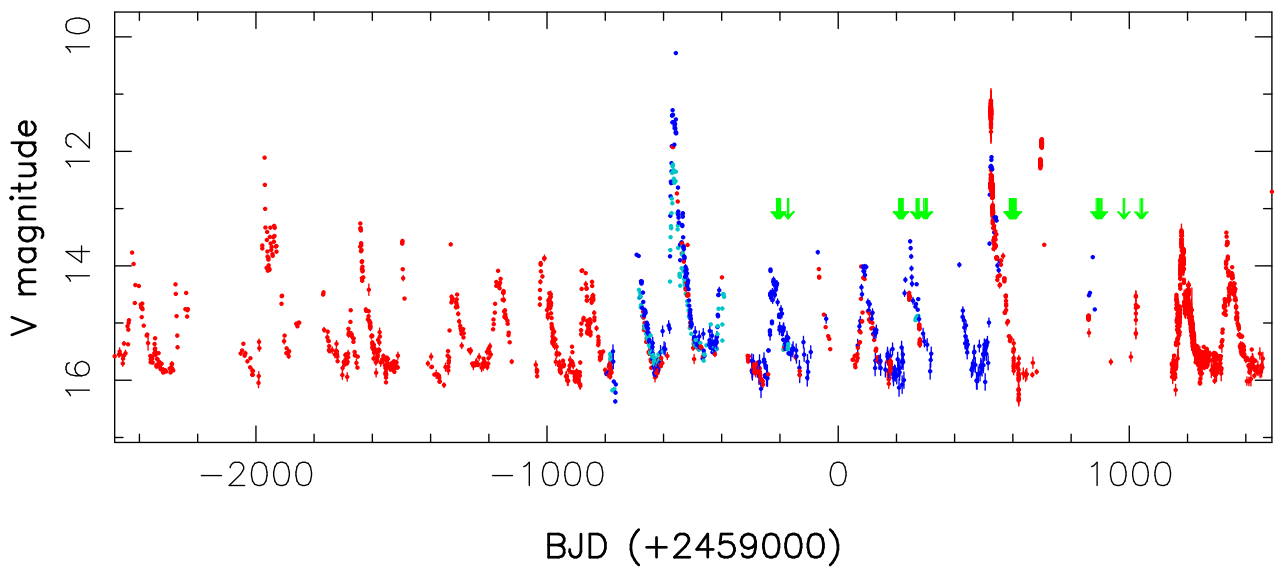}}
\caption[]{AAVSO, ASAS-SN and ZTF photometry of V347~Aur over the last 11~yr.  Measurements in the $V$ (AAVSO, ASAS-SN), $G$ (ASAS-SN, ZTF) and $R$ (ZTF) bands are depicted as 
red, blue and cyan dots (with $G$ and $R$ band data corrected by $-1.0$ and $+1.05$~mag respectively to match $V$ band data). BJDs of our SPIRou spectra are depicted with 
green arrows. The periodogram of these photometric data (not shown) peaks at 154.7~d, fully consistent with our estimate of the orbital period (see Table~\ref{tab:pla}). }  
\label{fig:pho}
\end{figure*}

\begin{figure}
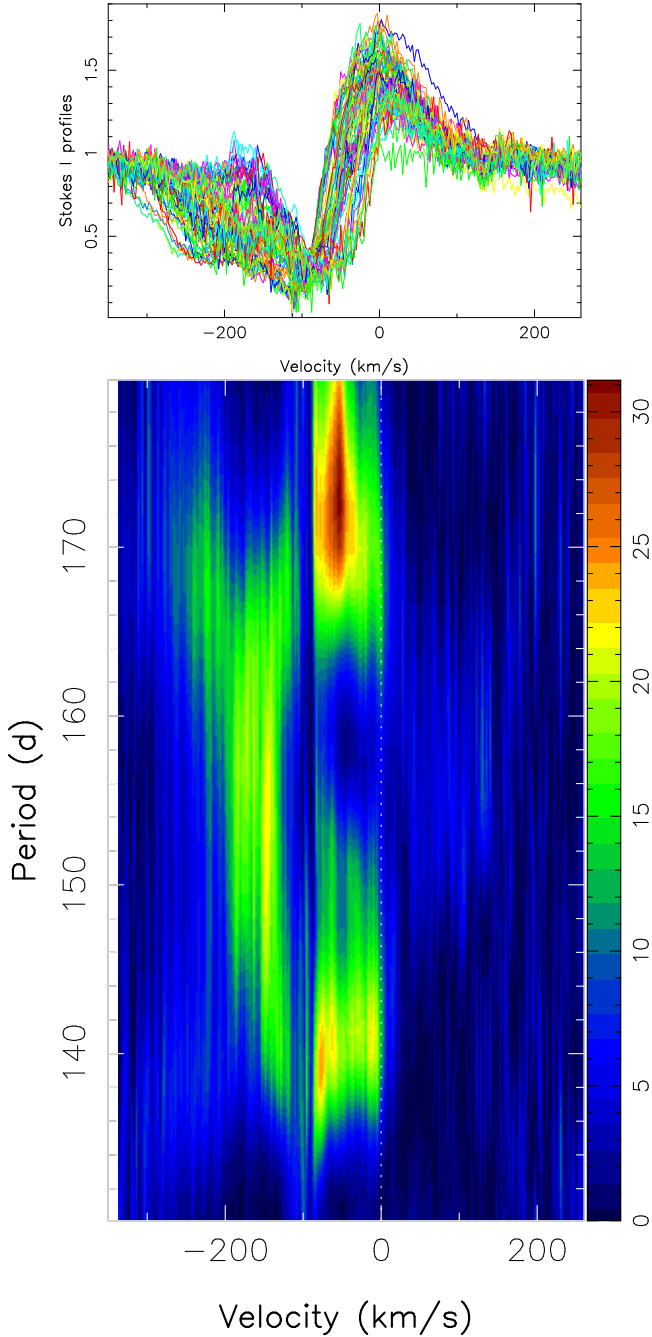

\centerline{\hspace{-2mm}\includegraphics[scale=0.3,angle=-90]{fig/v347aur-hei.ps}} 
\centerline{\includegraphics[scale=0.55,angle=-90]{fig/v347aur-hei-p2d.ps}}
\caption[]{Same as Fig.~\ref{fig:eml1} for the 1083.3-nm \hei\ triplet} 
\label{fig:eml2}
\end{figure}

\begin{figure}
\centerline{\includegraphics[scale=0.35,angle=-90]{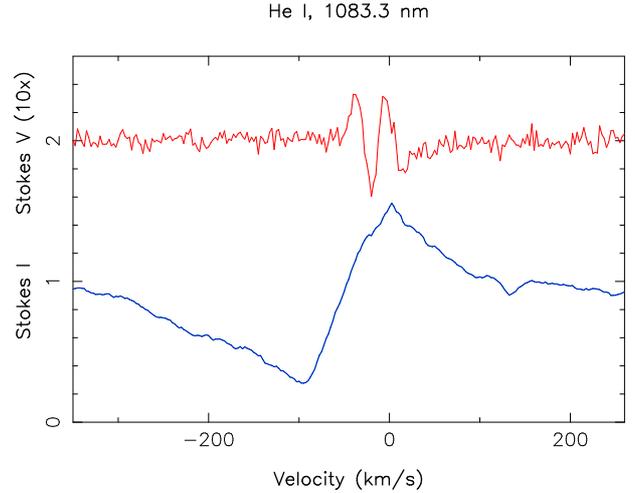}}
\caption[]{Weighted-average Stokes $I$ (blue) and $V$ (red) LSD profiles of the 1083.3-nm \hei\ line of V347~Aur over our full data set. 
Zeeman signatures are clearly detected in conjunction with the emission peak.  The Stokes $V$ LSD profile is expanded by a factor of 10$\times$ 
and shifted upwards by 2.  } 
\label{fig:emv}
\end{figure}

Both \pab\ and \brg\ are in emission at all times (see upper panels of Fig.~\ref{fig:eml1}).  Before scaling with veiling, EWs vary by a factor of 4--5$\times$ throughout 
our monitoring campaign, scale moderately well with $r_{JH}$ and $r_K$ ($R\simeq0.5$) and range from 61 to 249~\kms\ with a median of 130~\kms\ (0.26 to 1.07~nm, 
median 0.56~nm) for \pab\ and from 14 to 58~\kms\ with a median of 31~\kms\ (0.10 to 0.42~nm, median 0.22~nm) for \brg\ (see Table~\ref{tab:log}).  Once scaled up with 
veiling, EWs reach up to 441 and 254~\kms\ (1.89 and 1.84~nm) for \pab\ and \brg\ respectively, and their temporal variations, mimicking those of $r_{JH}$ and $r_K$, 
exhibit irregular peaks every $157\pm7$~d (again consistent with \Porb) coinciding with periastron passage.  The 2D periodograms of both lines, shown in the lower panels 
of Fig.~\ref{fig:eml1}, exhibit clear peaks at \Porb\ and at the first and second harmonics (not shown).  The periodograms also indicate that \pab\ and \brg\ vary as a 
whole with \Porb, with no obvious substructure exhibiting a different pattern than their EWs.  
In particular, we see no signal at \Prot\ in either lines that remains consistent over the full duration of our observations, and especially no red-shifted absorption 
events probing disk material accreted along magnetic funnels to high-latitude regions at the surface of the star (as in prototypical classical T~Tauri stars).  

{\emr To derive logarithmic mass accretion rates $\log\Mdot$, we scaled the veiling corrected EWs of \pab\ and \brg\ by the theoretical flux continuum of the appropriate 
stellar model to obtain line fluxes (in \lsun), then used the tabulated relations of \citet{Alcala17} to translate these lines fluxes into accretion luminosities, and 
finally applied the analytical relation of \citet{Gullbring98} giving mass accretion rates as a function of accretion luminosities and stellar parameters.  } 
The $\log\Mdot$ values that we obtain range from 
$-8.8$ to $-8.0$ (median $-8.5$, in \mspy) for \pab\ and from $-9.2$ to $-7.8$ (median $-8.7$) for \brg, with both independent estimates agreeing well together.  The peak 
accretion rates we infer are much weaker than those quoted in \citet{Dahm20}, reflecting that the accretion outbursts we monitored during our 
campaign are far less extreme than the mammoth event at BJD~$\simeq2458435$ described by \citet{Dahm20}, with $V$ band photometry rising by up to $\simeq$5.5~mag.  Archival 
AAVSO\footnote{https://www.aavso.org/}, ASAS-SN \citep[][]{Kochanek17} and ZTF \citep[][]{Bellm19} photometry (see Fig.~\ref{fig:pho}) indeed confirms that this is the case, 
with $V$ photometry rising by less than $\simeq$1.5 magnitudes during our SPIRou 
observations (which missed the main brightness peak at BJD~$\simeq2459526$ in this time slot).  The brightest event simultaneously monitored with ASAS-SN and SPIRou occurred 
at the beginning of our time series, and reached a brightness of only 2.4~per cent that of the largest outburst at BJD~$\simeq2458435$.  Attributing these brightness variations to 
changes in the accretion luminosity and extrapolating from our $\log\Mdot$ measurements 
would imply a peak $\log\Mdot\simeq-6.2$ (in \mspy) for the strongest accretion events of V347~Aur, an estimate that is 
consistent with that of \citet{Dahm20}.  The lowest accretion rates we infer from our data set ($\log\Mdot\simeq-9.0$ in \mspy) correspond to orbital phases away from the 
periastron and to periods of weak (and sometimes even null) veiling, both suggesting that accretion outside of the pulsed events is quite low and therefore that the central 
regions of the accretion disc are likely depleted in V347~Aur compared to prototypical single PMS stars.  Besides, we find that the periodogram of the archival photometric 
data over the last 11~yr peaks at 154.7~d, in perfect agreement with our determination of the orbital period of V347~Aur (see Table~\ref{tab:pla}).  

We finally note that the 1083.3-nm \hei\ triplet shows a P~Cygni profile with a blue-shifted absorption component extending to $-300$~\kms\ (and sometimes even $-400$~\kms) 
that suggests the presence of a strong stellar wind, rather than that of a conical disc wind which would be unable to intersect our line of sight to the star \citep[given 
the inclination angle $i$ and assuming a disc aligned with the equatorial plane of the star,][]{Kwan07}.  The \hei\ profile also features a slightly red-shifted emission peak 
varying in strength by over an order of magnitude (see top panel of Fig.~\ref{fig:eml2}) and extending up to about 100~\kms, probing ongoing accretion.  The 2D periodogram, 
depicted in the bottom panel of Fig.~\ref{fig:eml2} indicates that only the bluest wing (beyond --100~\kms) is modulated with the orbital period, as a likely response of the 
accretion-powered stellar wind to the QP accretion events.  Although not the strongest signal in the 2D periodogram, this modulation is nonetheless clearly detected, reaching 
powers similar to those associated with the \pab\ and \brg\ emission peaks in their 2D periodograms (see Fig.~\ref{fig:eml1}).  Surprisingly, the emission component redward of 
the line centre is not periodically modulated, even at shorter periods, suggesting longitudinally distributed (rather than localised) accretion, whereas the sharp transition between 
the absorption and emission components (between --100 and the line centre) is modulated with a longer period of $\simeq$173~d for unclear reasons (the weaker feature at 140~d 
being an alias of the main one).  As for \pab\ and \brg, we see no evidence of a consistent signal at \Prot, be it in the blue-shifted absorption or the in the emission components.  
We finally note that the median \hei\ profile shows a conspicuous Zeeman signature slightly blue shifted with respect to the line centre (see Fig.~\ref{fig:emv}), confirming that 
some of the emission comes from accretion regions at the stellar surface, and that the axisymmetric component of the magnetic topology in these accretion regions is complex 
\citep[rather than unipolar, as in, e.g.,TW~Hya,][]{Donati24b}.

\section{Summary and discussion}
\label{sec:dis}

We carried out a spectropolarimetric and velocimetric monitoring study of the $\simeq$0.8~Myr low-mass protostar V347~Aur ($\mstar=0.33\pm0.05$~\msun, see Sec.~\ref{sec:par}) 
with SPIRou at CFHT in the framework of the SLS and SPICE Large Programs, and collected 79 validated nIR Stokes $I$ and $V$ spectra of this object over a time 
frame of 1258~d covering 4 main seasons (from 2019 October to 2023 April).  

We find that V347~Aur is a spectroscopic binary with a brown dwarf companion, orbiting on an eccentric trajectory of period $\Porb=154.6\pm0.7$~d and semi-major 
axis $a=0.39\pm0.02$~au, and generating an RV wobble of semi-amplitude $K=1.05\pm0.06$~\kms.  The corresponding minimum mass of the companion is equal to 
$M_b \sin i=12.7\pm0.7$~\mjup, implying a mass of $M_b=29.0\pm1.6$~\mjup\ assuming the companion orbits within the equatorial plane of the star (inclined at 
$i=26\pm5$\degr\ to the line of sight).  It classifies V347~Aur~b as a bona fide brown dwarf, making it a rare member of the brown dwarf desert of 
companions around low-mass stars \citep{Marcy00}, and in particular around M dwarfs for which there is relatively few close substellar companions \citep{Artigau21}.  
The derived eccentricity, equal to $e=0.28\pm0.12$, is reminiscent of that reported for warm Jupiters, which may 
result from gravitational interactions with accretion discs featuring inner cavities \citep{Debras21}.  No such inner cavity has yet been reported for V347~Aur, but 
the low accretion rates measured over most of the orbital motion suggest that there may indeed be one (see below).  As previously reported in \citet{Dahm20}, we 
observe enhanced accretion episodes on V347~Aur, which we find are occurring as the companion reaches periastron, likely triggering gravitational instabilities 
in the disc leading to pulsed accretion onto the primary star.  We speculate that this process may be auto-amplifying, with the eccentric motion of the 
companion progressively inducing a more pronounced cavity, which in turn further increases the eccentricity.  This scenario however requires an inner cavity to 
start with, possibly caused by early planet formation in the inner disc, with the nascent bodies being later destabilized by the migrating brown dwarf companion 
before being engulfed in the primary star.  

We clearly detect the large-scale magnetic field of V347~Aur in the recorded Stokes $V$ LSD profiles, and derive \Bl\ values of only a few tens of G, several times 
weaker than those of prototypical classical or weak-line T~Tauri stars \citep{Finociety21,Finociety23,Donati24,Donati24b}.  We also show that \Bl\ undergoes 
temporal fluctuations with a semi-amplitude of about 20~G and a period of $76.5\pm1.2$~d, consistent within the error bar with \Porb/2.  (The other option that 
these fluctuations actually occur with a periodicity of \Porb\ is found to be less likely.) 
Besides, we detect the rotation period of V347~Aur in both \Bl\ and RV data, which we find to be equal to $4.12\pm0.03$~d.  
The corresponding rotational modulation of \Bl\ is only about 10~G, i.e., significantly smaller that the average \Bl\ value of about --50~G over most of our 
observations, indicating a large-scale magnetic topology that is mostly axisymmetric about the rotation axis.  On a longer time scale, we also see evolution of the average \Bl, rising 
to about --20~G in the last observing season, and suggesting that the overall large-scale field of V347~Aur, and in particular its poloidal component that \Bl\ is 
mostly sensitive to, may switch polarity in the forthcoming seasons.  

Applying ZDI to our Stokes $V$ LSD profiles, we find that the large-scale field of V347~Aur is mostly toroidal, with the poloidal component including only about 20~per cent 
of the reconstructed magnetic energy.  This poloidal component, largely axisymmetric for most of our observations, includes a dominant dipole field of about 30~G 
tilted at about $20\pm10$\degr\ to the rotation axis, which evolves into a non-axisymmetric poloidal component with a highly tilted dipole towards the last observing 
season.  

{\emr Moreover, we report a polarity switch of the toroidal component, changing from positive polarity before periastron in 2021~Jan to negative 
polarity after periastron in 2021~Feb.  We suspect that such polarity switches of the toroidal field happen twice per orbital cycle, once close to periastron 
and a second time around apoastron, leading to the other observed (negative to positive) polarity switches between 2019~Dec and 2020~Dec, and between 2022~Nov and 2023~Feb.} 
Surprisingly, the poloidal component keeps the same polarity over the whole orbital cycle, but is likely modulated in amplitude with \Porb/2 (about a non-zero value) given the 76.5-d 
fluctuation observed for \Bl\ {\emr (mostly sensitive to the poloidal field)}.  This result comes as the first unambiguous observational evidence that accretion is able to impact 
magnetic field amplification and dynamo processes in low-mass PMS stars.  We speculate that the large-scale field of V347~Aur is (at least partly) {\emr induced   
by what we dub a ``pulsed dynamo'' mechanism} \citep[\emr by analogy with pulsed accretion,][]{Teyssandier20}, i.e., indirectly driven by the orbital motion through the 
enhanced accretion events at periastron passage {\emr that act as a forcing mechanism}.  The accreted material and angular 
momentum from the disc, regularly injected into the top convective layers of the star, are indeed likely affecting the convective pattern, inducing enhanced radial and 
latitudinal shearing and thereby boosting the large-scale toroidal field before it relaxes away from periastron (on a convective turnover time).    
{\emr Detailed MHD simulations are obviously required to further investigate these ideas, and to explore in a more quantitative manner how the toroidal and poloidal 
components of the large-scale dynamo field are expected to respond with time to such a forcing mechanism}.  
Dynamo processes are also likely to operate on a longer timescale in V347~Aur, with the poloidal component evolving from a mainly axisymmetric 
to a mainly non-axisymmetric configuration in the last season of our monitoring campaign.  

The large-scale field of V347~Aur shares similarities with that of the classical T~Tauri star V2247~Oph, of nearly equal mass (0.35~\msun), featuring a complex and 
dominant toroidal field \citep{Donati10}.  V347~Aur is to date the most extreme such example, with a large-scale field strength weaker than 100~G and a toroidal field 
which stores about 80~per cent of the magnetic energy during most of our observations.  This is consistent with our results and those of \citet{Flores19} regarding the 
small-scale field of V347~Aur, also found to be weaker than that of more massive prototypical T~Tauri stars \citep[e.g.,][]{Johns07,Lopez-Valdivia21}.  Besides, the 
large-scale field of V347~Aur shares similarities with that of the (more massive) classical T~Tauri eccentric close 
binary DQ~Tau \citep[of mass 0.6~\msun,][]{Pouilly23,Pouilly24}, both featuring a strong toroidal component on the primary star, but exhibiting significant differences 
as well.  Not only is the large-scale field much weaker in V347~Aur, but \Bl\ and the reconstructed toroidal field are both modulated with \Porb\ or its first harmonic 
in V347~Aur, whereas no such modulation is observed in DQ~Tau despite the latter being more eccentric ($e=0.59$).  We speculate that this comes from \Porb\ being much 
longer in V347~Aur and comparable to the convective turnover time in very-low-mass stars \citep[e.g.,][]{Wright18}, thereby leaving time for dynamo processes to react 
to the varying accretion conditions across the orbit and to reflect them in the large-scale field, whereas dynamo rather averages out the rapidly varying accretion 
pattern of DQ~Tau.  

From the veiling in $JH$ and $K$ bands, we confirm that accretion strengthens near periastron passage in V347~Aur.  Given the measured eccentricity ($e=0.28\pm0.12$) 
and semi-major axis ($a=0.39\pm0.02$~au) causing the brown dwarf companion to orbit at distances varying from $a_{\rm p}=0.28\pm0.05$~au at periastron to 
$a_{\rm a}=0.50\pm0.05$~au at apoastron, i.e., not as close to the primary star as in DQ~Tau or HD~104237, we speculate following \citet{Dahm20} that enhanced accretion 
occurs as a result of disk instabilities triggered near periastron passage by the brown dwarf companion.  Accretion bursts in V347~Aur are known to induce brightness 
variations of up to 5.5~mag in $V$ \citep[][see also Fig.~\ref{fig:pho}]{Dahm20}.  From archival AAVSO, ASAS-SN and ZTF photometry, we find that accretion outbursts 
monitored during our monitoring campaign are weak ones, with brightness rising by $\Delta V<1.5$~mag.  
We also report that the \pab\ and \brg\ lines, well-known accretion proxies, also show clear emission peaks near periastron passage.  From \pab\ and \brg\ emission 
fluxes corrected from veiling, and using the scaling relations of \citet{Alcala17}, we infer logarithmic mass accretion rates (in \mspy) at the time of our observations 
ranging from about --9.0 away from periastron up to about --7.8 at periastron passage in V347~Aur.  Extrapolating to the strongest outburst recorded on V347~Aur, 
we obtain peak logarithmic mass accretion rates of $-6.2$ \citep[in \mspy, consistent with, the results of][]{Dahm20}.   
The low accretion rates away from periastron also suggest that the inner accretion disc of V347~Aur is significantly depleted with respect to prototypical 
accretion discs of protostars, and may have acted as a cavity causing the orbital motion of the brown dwarf companion to become eccentric \citep{Debras21}.  
We suggest that V347~Aur is a low-mass version of the eccentric binary HD~104237, where the system is apparently able to clear out a large 
central cavity in the disc and to give rise to complex accretion streams bridging the disc and the stars, with accretion rates varying by orders of magnitude 
\citep{Dunhill15}.  This is qualitatively similar to what we see on V347~Aur.   

Following \citet{Bessolaz08}, we infer that the weak large-scale magnetic dipole of V347~Aur is not able to disrupt the accretion disc beyond a radius of 
$\rmag\simeq1.3$~\rstar\ (i.e., $\rmag/\rcor\simeq0.4$), even for a logarithmic accretion rate as low as --9.0, whereas the accretion flow is likely to be barely 
impacted by the magnetic field in high / extreme accretion episodes.  We thus speculate that disk material destabilized by the companion at periastron passage is 
likely to be more or less free-falling through the field lines towards equatorial regions at the surface of the star, and that accretion will proceed in an 
unstable fashion at the lowest accretion rates away from periastron, via tongues of disk material penetrating the field \citep[as in][]{Blinova16}.  
We note that the 1083.3-nm \hei\ triplet is also modulated at \Porb, but only in its blue-shifted absorption component, suggesting that the star is triggering 
an accretion-powered wind getting stronger during accretion episodes near periastron.  On the contrary, the \hei\ emission component is not periodically modulated, 
neither at \Porb\ nor at \Prot, as a likely result of accretion taking place in a chaotic way over most stellar longitudes rather than on stable regions at the footpoints 
of accretion funnels, as in prototypical classical T~Tauri stars \citep[e.g.,][]{Donati24,Donati24b}.  The complex Zeeman signature detected in the median \hei\ 
profile is further evidence in this direction.  

More observations of V347~Aur with SPIRou such as those reported in this study would be quite worthwhile, gathered this time more evenly across the whole orbital 
motion and over several orbital cycles featuring accretion outbursts of variable strengths at periastron passages.  
Besides, high-angular resolution interferometric observations at mm (e.g., NOEMA) or IR (e.g., GRAVITY) wavelengths should come as a great addition to further 
constraint the disk properties, in particular its inclination and extent, as well as the dynamical mass of the system.  
Such data would allow us to refine the 
characterization of this very interesting and rather unique protostar for our understanding of star / planet formation, in particular in binary systems which account 
for a large fraction of the low-mass star population, and more specifically for unveiling how magnetic fields are able to impact this process.  
With its unexpectedly weak 
large-scale field, a likely result of the strong-accretion episodes modifying the convection pattern in the stellar interior and apparently triggering a ``pulsed-dynamo'' mechanism, 
V347~Aur is especially interesting in this respect and definitely deserves sustained interest from the young low-mass star community over the forthcoming 
years.

\section*{Acknowledgements}
We thank an anonymous referee for valuable comments which allowed us to improve the manuscript.  
This project received funds from the European Research Council (ERC) under the H2020 research \& innovation program (grant agreement \#740651 NewWorlds), 
the Agence Nationale pour la Recherche (ANR, project ANR-18-CE31-0019 SPlaSH) and the Investissements d'Avenir program (project ANR-15-IDEX-02).    
We thank Silvia Alencar and Hsien Shang for comments that helped clarify an earlier version of the manuscript.  
This work benefited from the SIMBAD CDS database at URL {\tt http://simbad.u-strasbg.fr/simbad}, the AAVSO database at URL {\tt https://www.aavso.org}  and the ADS system 
at URL {\tt https://ui.adsabs.harvard.edu}.
Our study is based on data obtained at the CFHT, operated by the CNRC (Canada), INSU/CNRS (France) and the University of Hawaii.
The authors wish to recognise and acknowledge the very significant cultural role and reverence that the summit of Maunakea has always had
within the indigenous Hawaiian community.  We are most fortunate to have the opportunity to conduct observations from this mountain.

\section*{Data availability}  SLS data are publicly available from the Canadian Astronomy Data Center, whereas SPICE data will be available at the same place 
from mid 2024 to mid 2025.

\bibliography{v347aur} 
\bibliographystyle{mnras}

\appendix

\section{Observation log}
\label{sec:appA}

Table~\ref{tab:log} gives the full log and associated \Bl\ and RV measurements at each observing epoch from our SPIRou spectra.

\begin{table*}
\small
\caption[]{Observing log of our SPIRou observations of {\emr V347~Aur} in seasons 2019, 2020, 2021 and 2022.  All exposures consist of 4 sub-exposures of equal length.
For each visit, we list the barycentric Julian date BJD, the UT date, the rotation cycle c and phase $\phi$ (computed as indicated in Sec.~\ref{sec:obs}),
the total observing time t$_{\rm exp}$, the peak SNR in the spectrum (in the $H$ band) per 2.3~\kms\ pixel, the noise level in the LSD Stokes $V$ profile,
the estimated \Bl\ with error bars, the nightly averaged RVs and corresponding error bars, the veiling in the $JH$ and in the $K$ bands $r_{JH}$ and $r_K$ 
measured from LSD profiles of atomic lines and CO lines respectively, and finally the EWs 
of the \pab\ and \brg\ emission lines with error bars (not scaled up with veiling).  }
\begin{tabular}{ccccccccccc}
\hline
BJD        & UT date & c / $\phi$ & t$_{\rm exp}$ & SNR & $\sigma_V$            & \Bl\   &  RV    & $r_{JH}$ / $r_K$ & EW \pab\ & EW \brg\ \\
(2459000+) &         &            &   (s)        & ($H$) & ($10^{-4} I_c$)       & (G)   & (\kms)  &                  & (\kms)   & (\kms)   \\
\hline
-211.9145299 & 31 Oct 2019 & 0 / 0.021 & 2407.1 & 245 & 2.43 & -38$\pm$21 & 9.22$\pm$0.28 & 0.77 / 1.75 & 249$\pm$10 & 58$\pm$3 \\
-210.9098887 & 01 Nov 2019 & 0 / 0.265 & 2407.1 & 236 & 2.49 & -32$\pm$21 & 8.62$\pm$0.29 & 0.75 / 2.54 & 247$\pm$10 & 57$\pm$3 \\
-208.9302687 & 03 Nov 2019 & 0 / 0.745 & 2407.1 & 230 & 2.54 & -76$\pm$20 & 8.65$\pm$0.29 & 0.57 / 3.40 & 224$\pm$10 & 54$\pm$3 \\
-206.8956725 & 05 Nov 2019 & 1 / 0.239 & 2407.1 & 219 & 2.69 & -26$\pm$20 & 8.95$\pm$0.29 & 0.53 / 2.12 & 218$\pm$10 & 56$\pm$3 \\
-204.9686594 & 07 Nov 2019 & 1 / 0.707 & 2407.1 & 213 & 2.77 & -60$\pm$20 & 9.02$\pm$0.29 & 0.46 / 2.28 & 209$\pm$10 & 51$\pm$3 \\
-202.9959549 & 09 Nov 2019 & 2 / 0.185 & 2407.1 & 149 & 4.21 & -19$\pm$30 & 7.96$\pm$0.35 & 0.45 / 1.16 & 202$\pm$10 & 52$\pm$3 \\
-200.9071390 & 11 Nov 2019 & 2 / 0.692 & 2407.1 & 172 & 3.67 & -66$\pm$22 & 8.84$\pm$0.24 & 0.25 / 0.84 & 151$\pm$10 & 39$\pm$3 \\
-198.9782271 & 13 Nov 2019 & 3 / 0.161 & 2407.1 & 212 & 2.92 & -65$\pm$19 & 8.77$\pm$0.31 & 0.34 / 1.29 & 190$\pm$10 & 50$\pm$3 \\
-197.9612905 & 14 Nov 2019 & 3 / 0.407 & 2407.1 & 190 & 3.16 & -43$\pm$21 & 8.43$\pm$0.31 & 0.34 / 1.03 & 154$\pm$10 & 42$\pm$3 \\
-174.0001745 & 08 Dec 2019 & 9 / 0.223 & 2407.1 & 182 & 3.26 & -34$\pm$19 & 8.10$\pm$0.27 & 0.20 / 0.71 & 134$\pm$10 & 32$\pm$3 \\
-172.9716427 & 09 Dec 2019 & 9 / 0.473 & 2407.1 & 195 & 3.02 & -73$\pm$17 & 8.39$\pm$0.25 & 0.17 / 0.69 & 124$\pm$10 & 29$\pm$3 \\
-172.0535410 & 10 Dec 2019 & 9 / 0.696 & 2312.3 & 214 & 2.71 & -40$\pm$15 & 8.44$\pm$0.20 & 0.15 / 0.68 & 129$\pm$10 & 29$\pm$3 \\
-170.9462697 & 11 Dec 2019 & 9 / 0.964 & 2407.1 & 210 & 2.76 & -36$\pm$16 & 8.68$\pm$0.25 & 0.18 / 0.69 & 123$\pm$10 & 29$\pm$3 \\
-169.9334152 & 12 Dec 2019 & 10 / 0.210 & 2379.2 & 192 & 3.07 & -30$\pm$17 & 8.21$\pm$0.20 & 0.14 / 0.65 & 115$\pm$10 & 28$\pm$3 \\
\hline
207.9252214 & 24 Dec 2020 & 101 / 0.924 & 2407.1 & 259 & 2.31 & -19$\pm$15 & 7.57$\pm$0.19 & 0.29 / 0.81 & 88$\pm$10 & 22$\pm$3 \\
208.9343916 & 25 Dec 2020 & 102 / 0.169 & 2407.1 & 254 & 2.39 & -42$\pm$15 & 7.44$\pm$0.20 & 0.27 / 0.75 & 111$\pm$10 & 25$\pm$3 \\
209.8880836 & 26 Dec 2020 & 102 / 0.400 & 2407.1 & 252 & 2.60 & -12$\pm$16 & 7.13$\pm$0.20 & 0.24 / 0.55 & 97$\pm$10 & 23$\pm$3 \\
211.9500308 & 28 Dec 2020 & 102 / 0.900 & 2407.1 & 189 & 3.31 & -43$\pm$21 & 7.22$\pm$0.20 & 0.28 / 0.35 & 79$\pm$10 & 18$\pm$3 \\
212.9362811 & 29 Dec 2020 & 103 / 0.140 & 2407.1 & 224 & 2.70 & -45$\pm$17 & 7.30$\pm$0.21 & 0.26 / 0.32 & 103$\pm$10 & 29$\pm$3 \\
214.8980511 & 31 Dec 2020 & 103 / 0.616 & 2407.1 & 220 & 2.71 & -57$\pm$16 & 7.11$\pm$0.20 & 0.24 / 0.33 & 107$\pm$10 & 27$\pm$3 \\
215.8662680 & 01 Jan 2021 & 103 / 0.851 & 2407.1 & 165 & 3.76 & -68$\pm$25 & 7.24$\pm$0.22 & 0.37 / 0.55 & 100$\pm$10 & 26$\pm$3 \\
217.8901191 & 03 Jan 2021 & 104 / 0.342 & 2407.1 & 248 & 2.27 & -75$\pm$15 & 7.06$\pm$0.22 & 0.36 / 0.43 & 100$\pm$10 & 25$\pm$3 \\
218.8741188 & 04 Jan 2021 & 104 / 0.581 & 2407.1 & 250 & 2.29 & -59$\pm$14 & 6.81$\pm$0.19 & 0.26 / 0.40 & 90$\pm$10 & 20$\pm$3 \\
219.9045892 & 05 Jan 2021 & 104 / 0.831 & 2407.1 & 246 & 2.28 & -47$\pm$15 & 7.33$\pm$0.20 & 0.33 / 0.59 & 80$\pm$10 & 18$\pm$3 \\
220.8613214 & 06 Jan 2021 & 105 / 0.063 & 2407.1 & 254 & 2.19 & -68$\pm$15 & 7.38$\pm$0.22 & 0.43 / 0.77 & 107$\pm$10 & 26$\pm$3 \\
221.9203503 & 07 Jan 2021 & 105 / 0.320 & 2407.1 & 240 & 2.39 & -34$\pm$17 & 6.76$\pm$0.22 & 0.46 / 0.73 & 119$\pm$10 & 30$\pm$3 \\
222.9318705 & 08 Jan 2021 & 105 / 0.566 & 2407.1 & 241 & 2.34 & -73$\pm$16 & 6.87$\pm$0.21 & 0.36 / 0.76 & 101$\pm$10 & 23$\pm$3 \\
267.7272476 & 22 Feb 2021 & 116 / 0.439 & 2407.1 & 271 & 2.10 & -42$\pm$14 & 9.47$\pm$0.21 & 0.34 / 2.38 & 173$\pm$10 & 39$\pm$3 \\
268.8634318 & 23 Feb 2021 & 116 / 0.714 & 2407.1 & 262 & 2.16 & -30$\pm$14 & 9.51$\pm$0.25 & 0.30 / 2.45 & 191$\pm$10 & 42$\pm$3 \\
271.7845315 & 26 Feb 2021 & 117 / 0.423 & 2407.1 & 224 & 2.52 & -10$\pm$16 & 8.78$\pm$0.20 & 0.30 / 1.42 & 177$\pm$10 & 41$\pm$3 \\
273.8651093 & 28 Feb 2021 & 117 / 0.928 & 2407.1 & 182 & 3.27 & -38$\pm$21 & 9.05$\pm$0.22 & 0.31 / 1.43 & 177$\pm$10 & 40$\pm$3 \\
276.7836915 & 03 Mar 2021 & 118 / 0.637 & 2407.1 & 263 & 2.06 & -26$\pm$12 & 8.98$\pm$0.19 & 0.21 / 1.22 & 162$\pm$10 & 36$\pm$3 \\
277.7285551 & 04 Mar 2021 & 118 / 0.866 & 2407.1 & 255 & 2.10 & -55$\pm$12 & 9.03$\pm$0.19 & 0.17 / 1.09 & 162$\pm$10 & 37$\pm$3 \\
293.7342341 & 20 Mar 2021 & 122 / 0.751 & 2407.1 & 232 & 2.44 & -69$\pm$12 & 8.58$\pm$0.17 & 0.02 / 0.36 & 146$\pm$10 & 39$\pm$3 \\
297.7716803 & 24 Mar 2021 & 123 / 0.731 & 2407.1 & 221 & 2.53 & -51$\pm$13 & 8.48$\pm$0.17 & 0.03 / 0.33 & 116$\pm$10 & 29$\pm$3 \\
299.7983781 & 26 Mar 2021 & 124 / 0.223 & 2407.1 & 186 & 3.25 & -54$\pm$17 & 8.05$\pm$0.18 & 0.04 / 0.29 & 106$\pm$10 & 28$\pm$3 \\
300.7513668 & 27 Mar 2021 & 124 / 0.454 & 2407.1 & 206 & 3.02 & -35$\pm$15 & 8.14$\pm$0.17 & 0.04 / 0.37 & 109$\pm$10 & 27$\pm$3 \\
301.7807176 & 28 Mar 2021 & 124 / 0.704 & 2407.1 & 159 & 3.80 & -36$\pm$21 & 8.30$\pm$0.19 & 0.15 / 0.86 & 95$\pm$10 & 25$\pm$3 \\
304.7728769 & 31 Mar 2021 & 125 / 0.430 & 2407.1 & 223 & 2.63 & -79$\pm$13 & 7.95$\pm$0.17 & 0.01 / 0.44 & 102$\pm$10 & 22$\pm$3 \\
305.7334444 & 01 Apr 2021 & 125 / 0.663 & 2407.1 & 226 & 2.46 & -68$\pm$12 & 8.30$\pm$0.16 & 0.02 / 0.37 & 96$\pm$10 & 24$\pm$3 \\
\hline
587.9333571 & 08 Jan 2022 & 194 / 0.159 & 2407.1 & 273 & 2.13 & -59$\pm$15 & 9.08$\pm$0.19 & 0.44 / 1.63 & 177$\pm$10 & 41$\pm$3 \\
588.9350942 & 09 Jan 2022 & 194 / 0.402 & 2407.1 & 272 & 2.19 & -30$\pm$15 & 8.91$\pm$0.18 & 0.39 / 1.89 & 161$\pm$10 & 39$\pm$3 \\
589.9833304 & 10 Jan 2022 & 194 / 0.656 & 2407.1 & 282 & 2.30 & -1$\pm$15 & 9.24$\pm$0.19 & 0.36 / 1.66 & 143$\pm$10 & 35$\pm$3 \\
590.7944760 & 11 Jan 2022 & 194 / 0.853 & 2407.1 & 248 & 2.43 & -37$\pm$17 & 9.22$\pm$0.18 & 0.41 / 1.48 & 142$\pm$10 & 37$\pm$3 \\
591.9658790 & 12 Jan 2022 & 195 / 0.137 & 2407.1 & 250 & 2.31 & -44$\pm$16 & 9.44$\pm$0.19 & 0.37 / 1.18 & 175$\pm$10 & 46$\pm$3 \\
592.9683245 & 13 Jan 2022 & 195 / 0.381 & 2407.1 & 232 & 3.43 & -63$\pm$23 & 8.92$\pm$0.19 & 0.36 / 0.94 & 172$\pm$10 & 43$\pm$3 \\
593.8925375 & 14 Jan 2022 & 195 / 0.605 & 2407.1 & 223 & 3.15 & -40$\pm$21 & 8.93$\pm$0.18 & 0.33 / 0.93 & 165$\pm$10 & 42$\pm$3 \\
594.8359992 & 15 Jan 2022 & 195 / 0.834 & 2407.1 & 279 & 2.10 & -20$\pm$13 & 9.20$\pm$0.18 & 0.28 / 0.92 & 164$\pm$10 & 39$\pm$3 \\
595.9992366 & 16 Jan 2022 & 196 / 0.116 & 2407.1 & 244 & 2.54 & -80$\pm$16 & 8.94$\pm$0.20 & 0.29 / 0.92 & 172$\pm$10 & 43$\pm$3 \\
596.8879787 & 17 Jan 2022 & 196 / 0.332 & 2407.1 & 265 & 2.29 & -43$\pm$14 & 8.84$\pm$0.17 & 0.26 / 0.72 & 169$\pm$10 & 43$\pm$3 \\
597.9366485 & 18 Jan 2022 & 196 / 0.587 & 2407.1 & 271 & 2.47 & -62$\pm$15 & 8.76$\pm$0.18 & 0.26 / 0.76 & 169$\pm$10 & 39$\pm$3 \\
598.9194038 & 19 Jan 2022 & 196 / 0.825 & 2407.1 & 255 & 2.28 & -58$\pm$14 & 8.85$\pm$0.18 & 0.23 / 0.67 & 142$\pm$10 & 36$\pm$3 \\
599.9732831 & 20 Jan 2022 & 197 / 0.081 & 2407.1 & 255 & 2.42 & -73$\pm$15 & 8.35$\pm$0.19 & 0.25 / 0.83 & 147$\pm$10 & 38$\pm$3 \\
601.8978507 & 22 Jan 2022 & 197 / 0.548 & 2407.1 & 174 & 3.75 & -73$\pm$23 & 8.74$\pm$0.18 & 0.26 / 0.48 & 130$\pm$10 & 35$\pm$3 \\
602.9316132 & 23 Jan 2022 & 197 / 0.799 & 2407.1 & 258 & 2.35 & -43$\pm$14 & 8.97$\pm$0.17 & 0.20 / 0.50 & 119$\pm$10 & 29$\pm$3 \\
603.9124578 & 24 Jan 2022 & 198 / 0.037 & 2407.1 & 272 & 2.41 & -73$\pm$15 & 8.58$\pm$0.21 & 0.25 / 0.67 & 121$\pm$10 & 29$\pm$3 \\
604.9164418 & 25 Jan 2022 & 198 / 0.281 & 2407.1 & 194 & 3.34 & -64$\pm$21 & 8.18$\pm$0.18 & 0.25 / 0.57 & 116$\pm$10 & 30$\pm$3 \\
606.9054391 & 27 Jan 2022 & 198 / 0.763 & 2407.1 & 225 & 2.64 & -49$\pm$16 & 8.49$\pm$0.17 & 0.23 / 0.54 & 122$\pm$10 & 36$\pm$3 \\
\hline
\end{tabular}
\label{tab:log}
\end{table*}

\setcounter{table}{0}
\begin{table*}
\caption[]{continued}
\begin{tabular}{ccccccccccc}
\hline
BJD        & UT date & c / $\phi$ & t$_{\rm exp}$ & SNR & $\sigma_V$            & \Bl\   &  RV    & $r_{JH}$ / $r_K$ & EW \pab\ & EW \brg\ \\
(2459000+) &         &            &   (s)        & ($H$) & ($10^{-4} I_c$)       & (G)   & (\kms)  &                  & (\kms)   & (\kms)   \\
\hline
607.8850302 & 28 Jan 2022 & 199 / 0.001 & 2407.1 & 232 & 2.45 & -44$\pm$15 & 8.57$\pm$0.17 & 0.22 / 0.40 & 127$\pm$10 & 30$\pm$3 \\
608.8852573 & 29 Jan 2022 & 199 / 0.244 & 2407.1 & 228 & 3.24 & -81$\pm$20 & 8.33$\pm$0.16 & 0.25 / 0.46 & 112$\pm$10 & 27$\pm$3 \\
609.8900384 & 30 Jan 2022 & 199 / 0.488 & 2407.1 & 248 & 2.33 & -79$\pm$14 & 8.15$\pm$0.18 & 0.24 / 0.58 & 106$\pm$10 & 26$\pm$3 \\
610.9655285 & 31 Jan 2022 & 199 / 0.749 & 2407.1 & 250 & 2.32 & -58$\pm$14 & 8.18$\pm$0.17 & 0.26 / 0.66 & 96$\pm$10 & 24$\pm$3 \\
\hline
886.1012388 & 02 Nov 2022 & 266 / 0.529 & 2407.1 & 293 & 1.88 & -19$\pm$12 & 9.24$\pm$0.20 & 0.27 / 0.92 & 159$\pm$10 & 39$\pm$3 \\
890.1199631 & 06 Nov 2022 & 267 / 0.505 & 2407.1 & 272 & 2.10 & -47$\pm$15 & 8.83$\pm$0.24 & 0.42 / 1.31 & 130$\pm$10 & 30$\pm$3 \\
891.1149740 & 07 Nov 2022 & 267 / 0.746 & 2407.1 & 287 & 2.04 & 10$\pm$13 & 8.91$\pm$0.24 & 0.33 / 1.36 & 148$\pm$10 & 33$\pm$3 \\
892.1127054 & 08 Nov 2022 & 267 / 0.989 & 2407.1 & 270 & 2.10 & 10$\pm$13 & 9.35$\pm$0.22 & 0.26 / 1.26 & 138$\pm$10 & 33$\pm$3 \\
893.1035281 & 09 Nov 2022 & 268 / 0.229 & 2407.1 & 239 & 2.29 & -4$\pm$14 & 8.81$\pm$0.22 & 0.26 / 0.83 & 160$\pm$10 & 35$\pm$3 \\
895.1312799 & 11 Nov 2022 & 268 / 0.721 & 2407.1 & 223 & 2.58 & -32$\pm$15 & 8.94$\pm$0.22 & 0.19 / 0.34 & 113$\pm$10 & 25$\pm$3 \\
896.0325625 & 12 Nov 2022 & 268 / 0.940 & 2407.1 & 269 & 2.03 & -14$\pm$12 & 9.33$\pm$0.20 & 0.18 / 0.66 & 132$\pm$10 & 31$\pm$3 \\
899.1357048 & 15 Nov 2022 & 269 / 0.693 & 2407.1 & 257 & 2.22 & -33$\pm$12 & 8.79$\pm$0.20 & 0.09 / 0.51 & 122$\pm$10 & 29$\pm$3 \\
902.1108546 & 18 Nov 2022 & 270 / 0.415 & 2407.1 & 254 & 2.27 & -44$\pm$12 & 8.53$\pm$0.19 & 0.05 / 0.31 & 114$\pm$10 & 27$\pm$3 \\
905.0987525 & 21 Nov 2022 & 271 / 0.140 & 2407.1 & 211 & 2.65 & -21$\pm$13 & 8.51$\pm$0.20 & 0.02 / 0.08 & 139$\pm$10 & 31$\pm$3 \\
906.1044391 & 22 Nov 2022 & 271 / 0.385 & 2407.1 & 230 & 2.41 & -65$\pm$12 & 8.23$\pm$0.18 & 0.00 / 0.13 & 112$\pm$10 & 25$\pm$3 \\
978.9367863 & 03 Feb 2023 & 289 / 0.062 & 2407.1 & 227 & 2.44 & -19$\pm$14 & 7.14$\pm$0.18 & 0.14 / 0.68 & 93$\pm$10 & 20$\pm$3 \\
980.9163238 & 05 Feb 2023 & 289 / 0.543 & 2407.1 & 176 & 3.82 & -18$\pm$23 & 6.77$\pm$0.20 & 0.22 / 0.68 & 73$\pm$10 & 18$\pm$3 \\
983.9060551 & 08 Feb 2023 & 290 / 0.268 & 2407.1 & 238 & 2.36 & -38$\pm$14 & 7.19$\pm$0.19 & 0.19 / 0.41 & 61$\pm$10 & 14$\pm$3 \\
1036.7643629 & 02 Apr 2023 & 303 / 0.098 & 2407.1 & 246 & 2.41 & 13$\pm$23 & 9.33$\pm$0.28 & 0.92 / 5.14 & 208$\pm$10 & 41$\pm$3 \\
1041.7274405 & 07 Apr 2023 & 304 / 0.303 & 2407.1 & 261 & 2.17 & 5$\pm$18 & 8.74$\pm$0.28 & 0.71 / 2.97 & 188$\pm$10 & 35$\pm$3 \\
1045.7255305 & 11 Apr 2023 & 305 / 0.273 & 2407.1 & 187 & 3.86 & 14$\pm$29 & 9.24$\pm$0.26 & 0.52 / 1.97 & 153$\pm$10 & 28$\pm$3 \\
\hline
\end{tabular}
\end{table*}

{\emr 
\section{Spectral fit and corner plot}
\label{sec:appB}

As an example, we show in Fig.~\ref{fig:spc} a small portion of our template SPIRou spectrum of V347~Aur, along with the best fit achieved with ZeeTurbo 
\citep{Cristofari23}, as described in Sec.~\ref{sec:obs}, along with the corresponding corner plot (see Fig.~\ref{fig:corp}).     

\begin{figure*}
\centerline{\includegraphics[scale=0.6,bb=20 40 900 610]{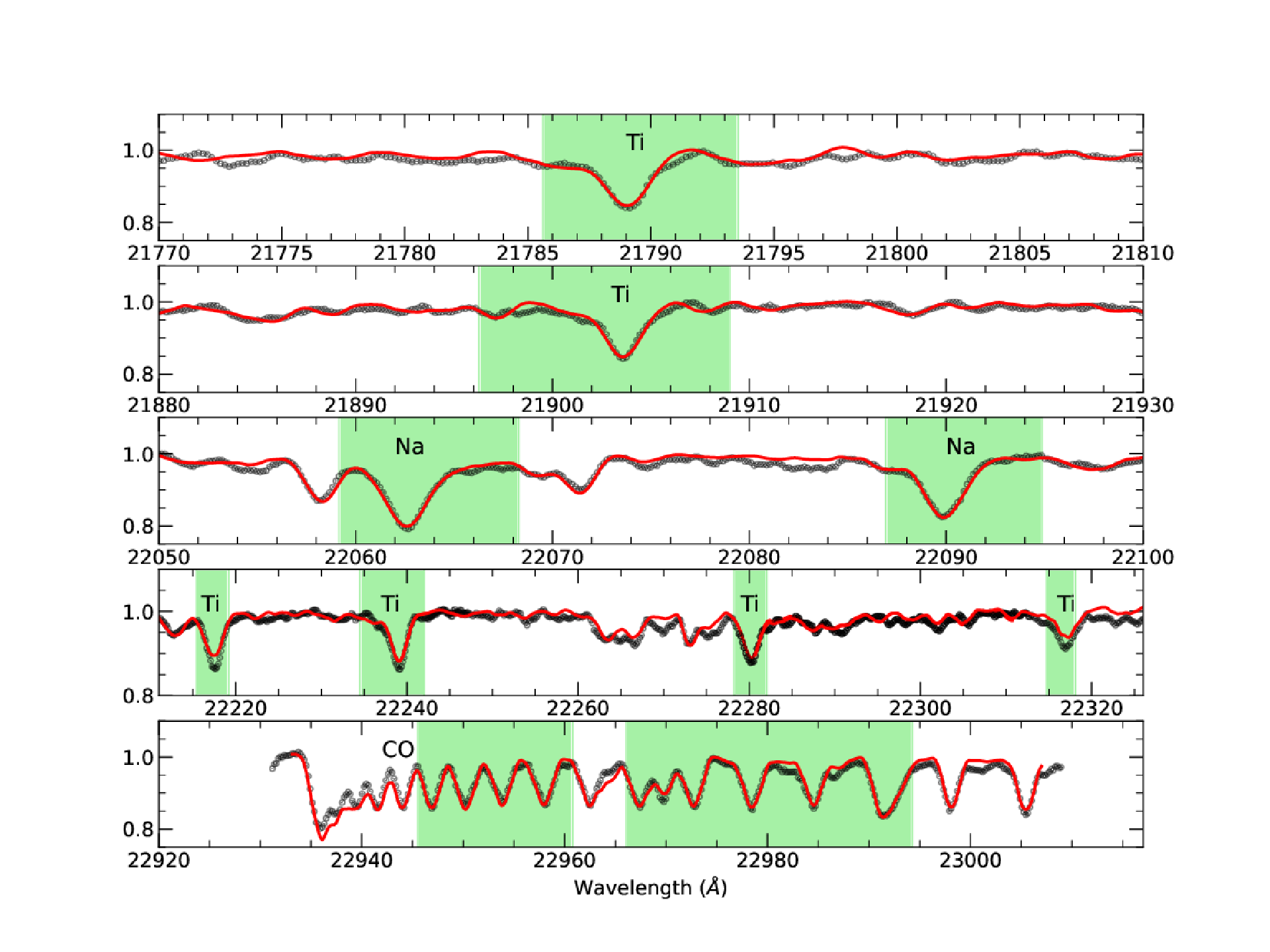}}
\caption[]{\emr Small portion of our template SPIRou spectrum of V347~Aur in the $K$ band (black circles), along with the optimal fit achieved 
with ZeeTurbo (red line) using the modeling approach of \citet{Cristofari23}.  The green areas indicate spectral regions considered in the 
fit. } 
\label{fig:spc}
\end{figure*}

\begin{figure*}
\centerline{\includegraphics[scale=0.53,bb=0 0 700 700]{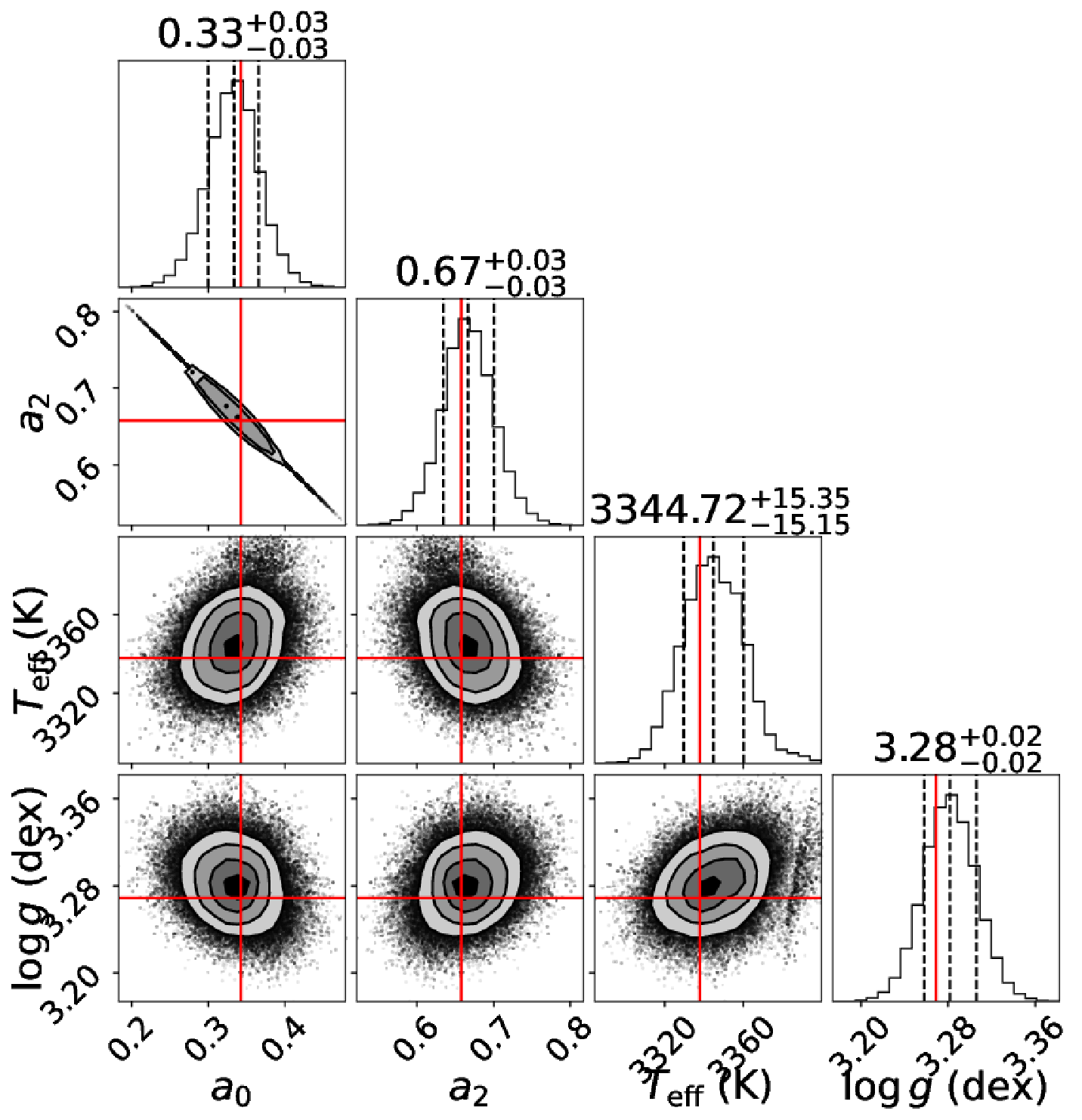}}
\caption[]{\emr Results of the MCMC fit to the template SPIRou spectrum of V347~Aur with ZeeTurbo.  We show here the corner plot and posterior 
distributions for the main variables, i.e., the 2 magnetic filling factors $a_0$ and $a_2$ corresponding to field strengths of 0 and 2~kG, 
\teff\ and \logg.  To account for additional systematic errors in the fitting process, we adopted conservative error bars for main stellar 
parameters \teff\ and \logg\ following \citet{Cristofari23} and rounded the derived estimates to $\teff=3340\pm50$~K and $\logg=3.30\pm0.10$.  } 
\label{fig:corp}
\end{figure*}

\section{Veiling in the $JH$ and $K$ bands}
\label{sec:appC}

We show in Fig.~\ref{fig:rjhk} the measured veiling of V347~Aur in the $JH$ and in the $K$ bands ($r_{JH}$ and $r_K$ respectively), plotted on the same graph as a function of time.  

\begin{figure}
\centerline{\includegraphics[scale=0.4,angle=-90]{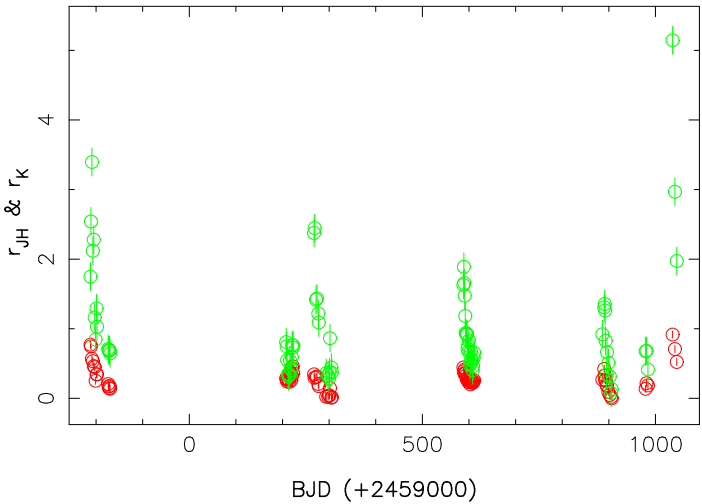}}
\caption[]{\emr Veiling of V347~Aur as a function of time, in the $JH$ and in the $K$ bands ($r_{JH}$ and $r_{K}$, red and green open circles respectively, with associated error bars), 
as measured from LSD profiles of atomic and CO bandhead lines. } 
\label{fig:rjhk}
\end{figure}
}

\bsp    
\label{lastpage}
\end{document}